\documentclass[aps,pre,groupedaddress]{revtex4}

\usepackage[dvipdfmx]{graphicx}
\usepackage{mathrsfs}
\usepackage{amsmath,amssymb}
\usepackage{hhline}
\usepackage{dcolumn}
\usepackage{bm}
\usepackage{here}
\usepackage{comment} 
\usepackage[dvips]{color}

\newcommand{\kr}{k_{\rm R}} 
\newcommand{\kb}{k_{\rm B}}
\newcommand{\vred}{v_{\rm R}}
\newcommand{\vb}{v_{\rm B}}

\newcommand{\tp}{\tilde{p}}
\newcommand{\tq}{\tilde{q}}

\newcommand{\bp}{\bar{p}}
\newcommand{\bq}{\bar{q}}
\newcommand{\br}{\bar{r}}
\newcommand{\bs}{\bar{s}}

\newcommand{\kmin}{k_{\rm min}}
\newcommand{\kmax}{k_{\rm max}}
\newcommand{\kcut}{k_{\rm cut}}

\newcommand{\tf}{\tilde{f}}
\newcommand{\df}{\Delta f}
\newcommand{\tfr}{\tilde{f}_{\rm R}}
\newcommand{\tfb}{\tilde{f}_{\rm B}}

\newcommand{\ered}{\epsilon_{\rm R}}
\newcommand{\eb}{\epsilon_{\rm B}}

\newcommand{\pr}{P_{\rm R}} 
\newcommand{\pb}{P_{\rm B}}

\newcommand{\xr}{x_{\rm R}} 
\newcommand{\xb}{x_{\rm B}}
\newcommand{\yr}{y_{\rm R}} 
\newcommand{\yb}{y_{\rm B}}

\newcommand{\qkk}{q_{k_{\rm R},k_{\rm B}}}
\newcommand{\rkk}{r_{k_{\rm R},k_{\rm B}}}

\newcommand{\qkkp}{q_{k_{\rm R}',k_{\rm B}'}}
\newcommand{\rkkp}{r_{k_{\rm R}',k_{\rm B}'}}

\newcommand{\be}{\begin{equation}}
\newcommand{\ee}{\end{equation}}
\newcommand{\ba}{\begin{eqnarray}}
\newcommand{\ea}{\end{eqnarray}}

\begin{document}

\title{Robustness of Random Networks with Selective Reinforcement against Attacks}
\author{Tomoyo Kawasumi}
\affiliation{Graduate School of Science and Engineering, Ibaraki University, 2-1-1, Bunkyo, Mito, 310-8512, Japan}
\author{Takehisa Hasegawa}
\email{takehisa.hasegawa.sci@vc.ibaraki.ac.jp}
\affiliation{Graduate School of Science and Engineering, Ibaraki University, 2-1-1, Bunkyo, Mito, 310-8512, Japan}

\begin{abstract}
We investigate the robustness of random networks reinforced by adding hidden edges against targeted attacks.
This study focuses on two types of reinforcement: uniform reinforcement, where edges are randomly added to all nodes, and selective reinforcement, where edges are randomly added only to the minimum degree nodes of the given network.
We use generating functions to derive the giant component size and the critical threshold for the targeted attacks on reinforced networks.
Applying our analysis and Monte Carlo simulations to the targeted attacks on scale-free networks, it becomes clear that selective reinforcement significantly improves the robustness of networks against the targeted attacks.
\end{abstract}

\maketitle

\section{Introduction} \label{Intro}

Networks, which consist of nodes and edges connecting nodes, have been studied extensively over the last few decades~\cite{albert2002statistical,newman2003structure,boccaletti2006complex}.
A remarkable feature of real networks, e.g., the World Wide Web, the Internet, social networking services, and air transportation networks, is that they are scale-free.
 In other words, their degree distribution $p(k)$ follows a power law, $p(k) \propto k^{-\gamma}$.
The heterogeneity of a network has a profound effect on its robustness against failures and attacks.
Albert et al. found that scale-free networks are considerably robust against random failures, where nodes are removed randomly from the network, but are highly vulnerable to targeted attacks, where nodes are removed in descending order of degree~\cite{albert2000error}.
This finding has been supported theoretically through percolation analysis of random networks~\cite{callaway2000network,cohen2000resilience,cohen2001breakdown}. 
Thus, the theory of network robustness has become a main pillar of network science, as noted by Gross and Barth in their review~\cite{gross2022network}.

Researchers in network science have explored two distinct approaches to improve the robustness of networks against attacks, i.e., network optimization and network reinforcement.
The optimization of a network topology is achieved by rewiring the edges under a fixed number of edges~\cite{liu2005optimization,xiao2010robustness,schneider2011mitigation,wu2011onion,herrmann2011onion,zhou2014memetic,park2016bypass,chujyo2021loop,lou2023structural}.
Schneider et al. demonstrated that an onion-like structure that is highly robust to attacks can be realized by repeating edge swaps so that the robustness measure $R$ increases~\cite{schneider2011mitigation}.
Chujyo and Hayashi focused on the loop structure of networks and demonstrated that the networks become robust by rewiring to enhance their loop structure~\cite{chujyo2021loop}.
The reinforcement of a network is achieved by adding edges to resist attacks~\cite{beygelzimer2005improving,zhao2009enhancing,li2012enhancing,ji2016improving,chan2016optimizing,cui2018enhancing,carchiolo2019network,kazawa2020effectiveness,dong2020improving,chen2022robustness,chujyo2022adding}.
Beygelzimer et al. numerically studied the effect of several different strategies comprising edge rewiring and edge addition on network robustness to demonstrate that preferential rewiring or addition between low degree nodes makes the network more robust than random ones~\cite{beygelzimer2005improving}.
Zhao et al. also demonstrated numerically that edge addition between low degree nodes weakens the degree heterogeneity of a network and makes it more robust to both failures and attacks~\cite{zhao2009enhancing}.
Some studies have further considered hiding additional edges from attacks as a practical strategy to protect scale-free networks~\cite{beygelzimer2005improving,zhuo2011improving,wang2021adversarial}.
For example, Zhuo et al.~\cite{zhuo2011improving} proposed an edge information protection strategy by adding hidden edges, and they demonstrated through simulations that hiding a small fraction of edges between low degree nodes contributes to the maintenance of the robustness of scale-free networks.
Thus, various simulations have suggested that the addition of (hidden) edges between low degree nodes is effective against attacks; however, analytical treatments for a comprehensive understanding of how to improve network robustness are completely lacking.

In this paper, we formulate the robustness of random networks reinforced by adding hidden edges against targeted attacks.
We focus on two types of network reinforcement: 
uniform reinforcement, where all nodes in a network are targeted for edge additions, and selective reinforcement, where selected nodes (minimum degree nodes in this study) are targeted for edge additions. 
We use generating functions to formulate the size of the giant component (GC) for the targeted attacks on the reinforced networks, thereby deriving the critical threshold.
We show that both uniform and selective reinforcement make scale-free networks robust. In particular, selective reinforcement significantly improves network robustness in terms of both the critical threshold and the robustness measure. 
Monte Carlo simulations confirm the validity of the theoretical treatments presented in this paper.

The remainder of this paper is organized as follows.
In Sec.~\ref{sec:analysis}, we formulate the robustness of reinforced networks against the targeted attacks using generating functions.
We derive the GC size under the attacks and the critical threshold for random networks with uniform reinforcement (Sec.~\ref{sec:uniform}) and selective reinforcement (Sec.~\ref{sec:selective}).
In Sec.~\ref{sec:result}, we confirm the validity of our analytical treatment by using Monte Carlo simulations, and we demonstrate that the selective reinforcement approach remarkably improves the robustness of scale-free networks against attacks.
Section~\ref{sec:summary} is devoted to the summary and discussion.

\begin{figure}
\centering
(a)
	\includegraphics[width=4.8cm]{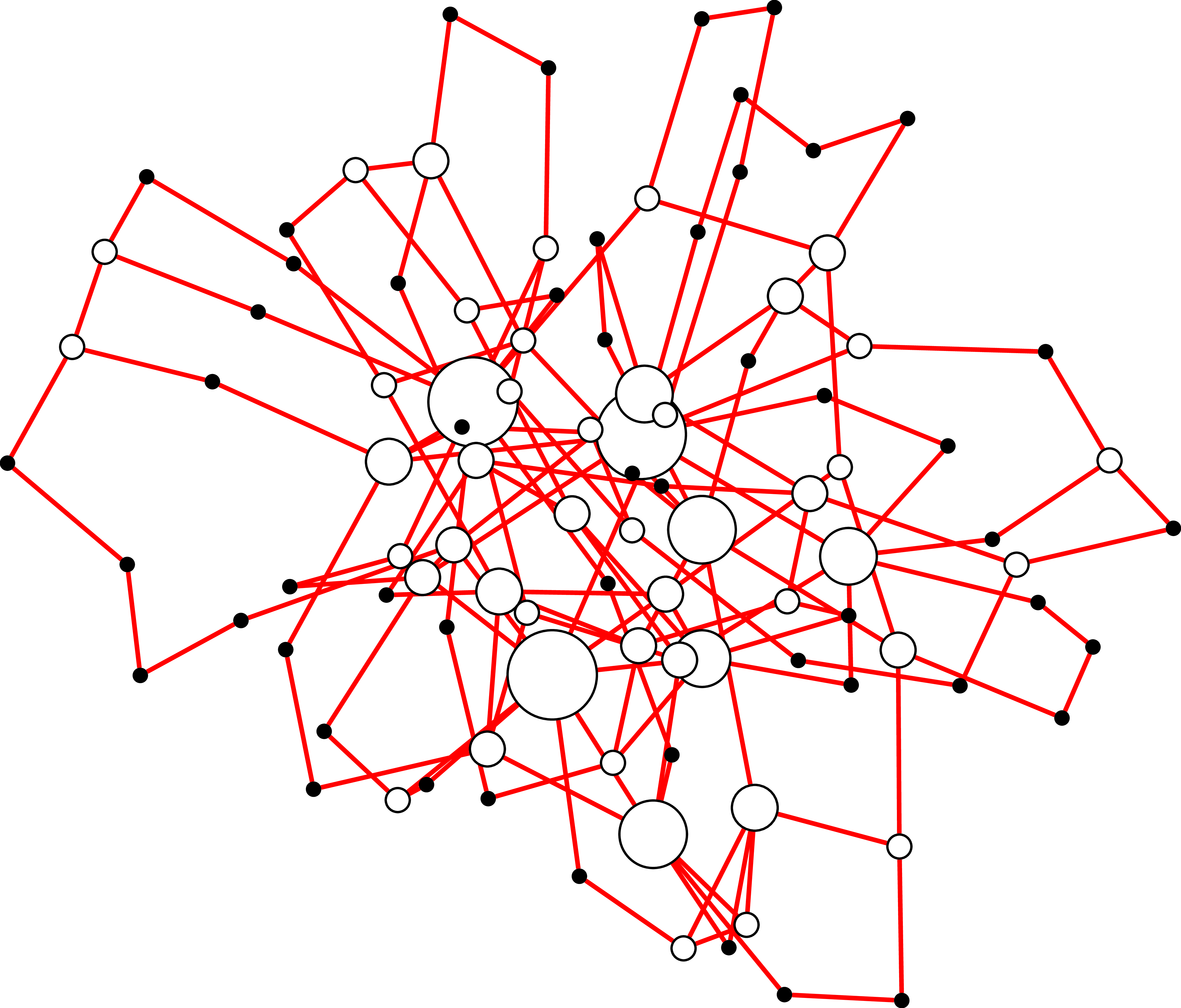}
(b)
	\includegraphics[width=5.2cm]{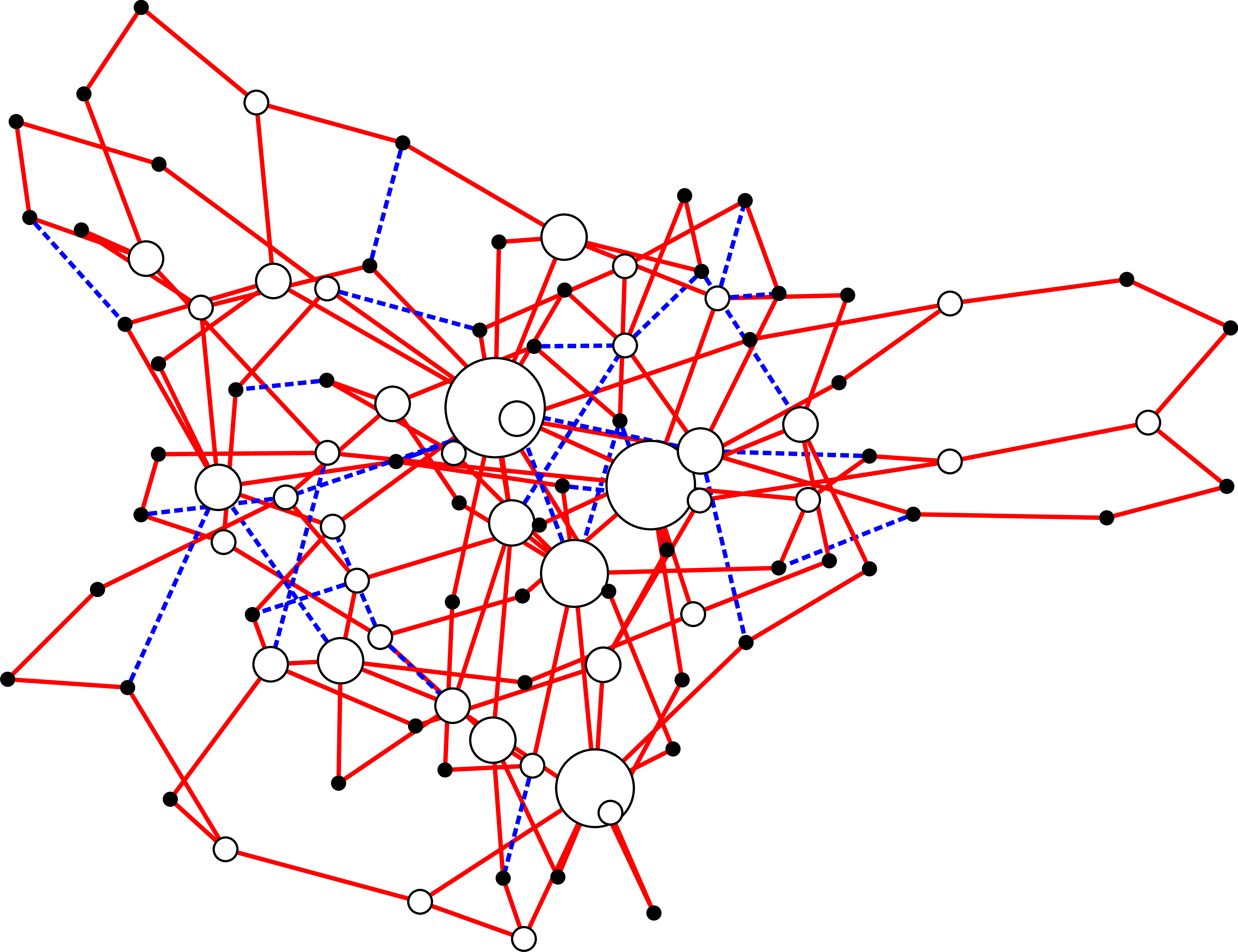}
(c)
	\includegraphics[width=5cm]{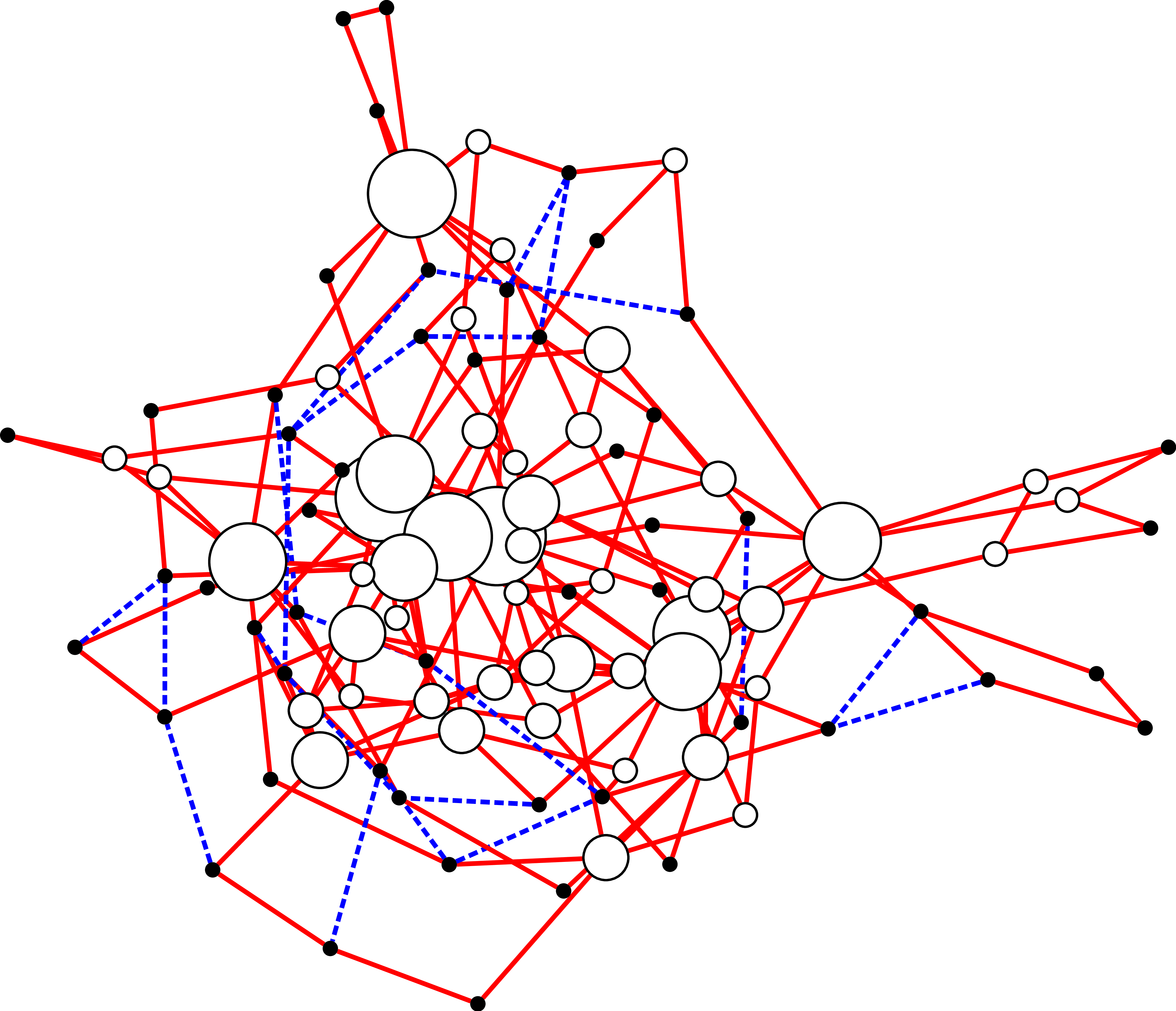}
\caption{
Examples of (a) a random network, (b) the network with uniform reinforcement, and (c) the network with selective reinforcement. 
Here, filled circles represent minimum (red) degree nodes and open circles represent all other nodes. 
Red solid lines represent red edges forming the original network and blue dotted lines represent blue edges added for reinforcement. 
In (c), blue edges are only connected to the minimum degree nodes.
}
\label{fig:schematic}
\end{figure}

\section{Formulation} \label{sec:analysis}

Let us assume an infinitely large and degree-uncorrelated network with degree distribution $p(k)$ ($k = \kmin, \kmin+1, \cdots, \kmax$).
Here $\kmin$ and $\kmax$ represent the minimum and maximum degree of the given network, respectively.
The reinforced network is obtained by adding hidden edges to the given network.
We refer to the edges forming the original network as {\it red edges} and the edges added for reinforcement as {\it blue edges}.
The number of red and blue edges connected to a node is termed {\it red degree} $\kr$ and {\it blue degree} $\kb$, respectively.
The present study focuses on {\it uniform reinforcement}, where all nodes are targeted for edge additions to be reinforced, and {\it selective reinforcement}, where minimum (red) degree nodes are targeted for edge additions and other nodes are not targeted.
Under the uniform reinforcement approach, blue edges can be added to all nodes; under the selective reinforcement approach, blue edges can be added only to minimum degree nodes (Fig.~\ref{fig:schematic}).
It should be noted that, under the selective reinforcement approach, a node connected by a blue edge always has a red degree of $k_{\rm min}$, i.e., $\kr = k_{\rm min}$.

The probability that a node selected at random from a network has red degree $\kr$ is $p(\kr)$.
When a red edge is randomly selected and a node is reached by traversing that edge, it will have $\kr+1$ red edges with probability $q(\kr)$, which is given as
\begin{equation}
q(\kr)=\frac{(\kr+1) p(\kr+1)}{\langle \kr \rangle} \quad {\rm where} \quad \langle \kr \rangle=\sum_{\kr} \kr p(\kr).
\end{equation}
We introduce the generating functions, $G_0(x)$ and $G_1(x)$, for $p(\kr)$ and $q(\kr)$ as 
\begin{equation}
G_0(x) = \sum_{\kr=\kmin}^{\kmax} p(\kr)x^{\kr} \quad {\rm and} \quad G_1(x) = \sum_{\kr=\kmin}^{\kmax} q(\kr)x^{\kr},
\end{equation}
respectively.

For additional edges, we denote by $\tp(\kb)$ the fraction of nodes that have blue degree $\kb$ to the nodes subject to reinforcement (all nodes in the uniform reinforcement case and the minimum degree nodes in the selective reinforcement case).
A node reached by traversing a randomly selected blue edge has $\kb + 1$ blue edges with probability $\tq(\kb)$, which is given as
\begin{equation}
\tq(\kb)=\frac{(\kb+1) \tp(\kb+1)}{\langle \kb \rangle} \quad {\rm where} \quad \langle \kb \rangle=\sum_{\kb} \kb \tp(\kb).
\end{equation}
We introduce the generating functions, $\tilde{G}_0(x)$ and $\tilde{G}_1(x)$, for $\tp(\kb)$ and $\tq(\kb)$ as 
\begin{equation}
\tilde{G}_0(x) = \sum_{\kb=0}^{\infty} \tp(\kb)x^{\kb} \quad {\rm and} \quad \tilde{G}_1(x) = \sum_{\kb=0}^{\infty} \tq(\kb)x^{\kb},
\end{equation}
respectively.

A targeted attack removes nodes in descending order of degree in the original network, i.e., in descending order of red degree. In this context, the blue edges are considered ``hidden from attacks,'' in the sense that the attack on nodes is determined solely based on their red edges, without taking their blue edges into account.
Let us consider an attack that deletes a fraction $1-f$ of nodes with the largest red degrees.
The maximum value of the red degree is reduced from $\kmax$ to a new cutoff degree $\kcut$.
All nodes with $\kmin \le \kr < \kcut$ are retained and those with $\kcut < \kr \le \kmax$ are deleted due to the attack.
For nodes with $\kr = \kcut$, a fraction $\Delta f$ ($1-\Delta f$) is randomly selected to be deleted (undeleted).
Then, the undeleted node fraction $f$ is related to $\kcut$ and $\Delta f$ as
\begin{equation}
f=\sum_{\kr=\kmin}^{\kcut} p(\kr) -\df p(\kcut) = \sum_{\kr = \kmin}^{\kcut; \Delta f} p(\kr). 
\label{def:f}
\end{equation}
Here, for convenience, we defined $\sum_{k = a}^{b; \Delta f} X(k) = \sum_{k = a}^{b} X(k) - \Delta f X(b)$.

We denote by $\tfr$ the probability that a node reached by following a red edge is not deleted.
Since a red edge is connected to a node of degree $\kr$ with probability $\kr p(\kr)/\langle \kr \rangle$,
\begin{equation}
\tfr = \sum_{\kr=\kmin}^{\kcut} \frac{\kr p(\kr)}{\langle \kr \rangle} -\Delta f \frac{\kcut p(\kcut)}{\langle \kr \rangle} = \sum_{\kr=\kmin}^{\kcut; \Delta f} \frac{\kr p(\kr)}{\langle \kr \rangle}.
\label{def:tfr}
\end{equation}
In addition, we denote by $\tfb$ the probability that a node reached by following a blue edge is not deleted.
The form of $\tfb$ depends on the reinforcement approach, as presented in the following subsections.

The reinforced network, which is obtained by adding hidden edges to the original network, is characterized by three types of joint probabilities: 
$p_{\kr, \kb}$, which is the probability that a node has $\kr$ red edges and $\kb$ blue edges;
$q_{\kr, \kb}$, which is the probability that a node reached by following a red edge has $\kr +1$ red edges and $\kb$ blue edges;
and $r_{\kr, \kb}$, which is the probability that a node reached by following a blue edge has $\kr$ red edges and $\kb+1$ blue edges.
To discuss the robustness of reinforced networks, we focus on the connected components of nodes that were not deleted by the attack.
After the attack, three probabilities, $p_{\kr, \kb}$, $q_{\kr, \kb}$, and $r_{\kr, \kb}$, change their shapes by the targeted removal of large degree nodes to new ones, $\bp_{\kr, \kb}$, $\bq_{\kr, \kb}$, and $\br_{\kr, \kb}$, respectively.
We introduce the generating functions for $\bp_{\kr, \kb}$, $\bq_{\kr, \kb}$, and $\br_{\kr, \kb}$ as 
\begin{subequations}
\begin{align}
F_p(x,y) &= \sum_{\kr=0}^{\kcut} \sum_{\kb=0}^\infty \bp_{\kr, \kb} x^{\kr} y^{\kb}, \\
F_q(x,y) &= \sum_{\kr=0}^{\kcut} \sum_{\kb=0}^\infty \bq_{\kr, \kb} x^{\kr} y^{\kb}, \\
\intertext{and}
F_r(x,y) &= \sum_{\kr=0}^{\kcut} \sum_{\kb=0}^\infty \br_{\kr, \kb} x^{\kr} y^{\kb},
\end{align}
\label{def:Fpqr}
\end{subequations}
respectively.

A GC, defined as the connected component that occupies a finite fraction of an infinitely large network, should form unless the network is decomposed.
The normalized GC size for undeleted nodes, $\bs_{\rm GC}$, which is equal to the probability that a randomly selected node is undeleted and belongs to the GC, is expressed as follows:
\begin{equation}
\bs_{\rm GC} = f (1 - F_p (\vred, \vb)),
\label{def:bs_GC}
\end{equation}
where $\vred$ is the probability that an undeleted node reached by following a red edge does not lead to the GC, and $\vb$ is the probability that an undeleted node reached by following a blue edge does not lead to the GC.
It was noted that $f$ represents the probability of a randomly selected node being undeleted and $\sum_{\kr=0}^{\kcut} \sum_{\kb=0}^\infty \bp_{\kr, \kb} \vred^{\kr} \vb^{\kb} = F_p(\vred, \vb)$ on the right hand side of Eq.~(\ref{def:bs_GC}) represents the probability of an undeleted node not being a member of the GC.
When the original network is degree-uncorrelated and additional edges are distributed in a random fashion, both $\vred$ and $\vb$ can be determined as the solution of the self-consistent equations,
\begin{equation}
\vred = F_q(\vred, \vb)
\quad
{\rm and}
\quad
\vb = F_r(\vred, \vb).
\label{def:vrvb}
\end{equation}

In the following subsections, we apply these formulations to random networks with the uniform reinforcement and selective reinforcement approaches.

\subsection{Uniform Reinforcement} \label{sec:uniform}

First, we consider the targeted attack on a network with uniform reinforcement, where the edges for reinforcement are added randomly to all nodes.
Under uniform reinforcement, each node is assigned the blue degree with probability $\tp(\kb)$ independent of its red degree.
The joint probabilities for the uniformly reinforced networks are as expressed follows:
\begin{equation}
p_{\kr, \kb} = p(\kr)\tp(\kb),
\quad
q_{\kr, \kb} = q(\kr)\tp(\kb),
\quad
r_{\kr, \kb} = p(\kr)\tq(\kb).
\end{equation}
Considering the targeted attack on the uniformly reinforced networks, we have 
\begin{equation}
\tfb= f, 
\end{equation} 
because blue edges are connected to nodes regardless of red edges at random and the probability that the red degree of the node reached by following a blue edge is $\kr$ is simply $p(\kr)$.

Let us derive the joint probability $\bp_{\kr, \kb}$ for undeleted nodes after the attack.
We begin with the conditional probability $P({\rm undel}, \kr, \kb | \kr', \kb')$ that a randomly chosen node is undeleted and has $\kr$ red edges and $\kb$ blue edges after the attack, given that it has $\kr'$ red edges and $\kb'$ blue edges prior to the attack.
Noting that a node is undeleted under the attack of $f$ only when the red degree is equal to or less than $\kcut$ and each neighbor connected to it via a red edge (blue edge) is undeleted with probability $\tfr$ ($\tfb$), this probability is given as
\begin{equation}
P({\rm undel}, \kr, \kb | \kr', \kb') = 
\begin{cases}
\binom{\kr'}{\kr} \tfr^{\kr} (1-\tfr)^{\kr'-\kr} \binom{\kb'}{\kb} \tfb^{\kb} (1-\tfb)^{\kb'-\kb} & (\kr' < \kcut) \\
(1-\df) \binom{\kr'}{\kr} \tfr^{\kr} (1-\tfr)^{\kr'-\kr} \binom{\kb'}{\kb} \tfb^{\kb} (1-\tfb)^{\kb'-\kb} & (\kr' = \kcut) \\
0 & (\kr' > \kcut)
\end{cases}.
\end{equation}
The probability $P({\rm undel}, \kr, \kb)$ that a node is undeleted and has $\kr$ red edges and $\kb$ blue edges connected to other undeleted nodes is
\begin{eqnarray}
P({\rm undel}, \kr, \kb) 
&=& \sum_{\kr'=\kr}^{\kmax} \sum_{\kb'=\kb}^{\infty} P({\rm undel}, \kr, \kb | \kr', \kb') p_{\kr, \kb} \nonumber \\
&=& \sum_{\kr'=\kr}^{\kcut; \df} p(\kr') \binom{\kr'}{\kr} \tfr^{\kr} (1-\tfr)^{\kr'-\kr} \sum_{\kb'=\kb}^{\infty} \tp(\kb') \binom{\kb'}{\kb} \tfb^{\kb} (1-\tfb)^{\kb'-\kb}.
\label{eq:Pundel}
\end{eqnarray}
Since the joint probability $\bp_{\kr, \kb}$ for undeleted nodes is defined as 
\begin{equation}
\bp_{\kr, \kb} = P({\rm undel}, \kr, \kb)/P({\rm undel}), \label{eq:jointPro}
\end{equation}
where $P({\rm undel})$ is the probability of a randomly selected node being undeleted, i.e., $P({\rm undel})=f$, we have
\begin{eqnarray}
\label{def:abp}
\bp_{\kr, \kb} = \frac{1}{f} \sum_{\kr'=\kr}^{\kcut; \df} p(\kr') \binom{\kr'}{\kr} \tfr^{\kr} (1-\tfr)^{\kr'-\kr} \sum_{\kb'=\kb}^{\infty} \tp(\kb') \binom{\kb'}{\kb} \tfb^{\kb} (1-\tfb)^{\kb'-\kb}.
\end{eqnarray}

Similarly, we obtain the probability $\bq_{\kr, \kb}$ that an undeleted node reached by following a red edge has $\kr+1$ red edges and $\kb$ blue edges after the attack as
\begin{eqnarray}
\label{def:abq}
\bq_{\kr, \kb} = \frac{1}{\tfr} \sum_{\kr'=\kr}^{\kcut-1; \df} q(\kr') \binom{\kr'}{\kr} \tfr^{\kr} (1-\tfr)^{\kr'-\kr} \sum_{\kb'=\kb}^{\infty} \tp(\kb') \binom{\kb'}{\kb} \tfb^{\kb} (1-\tfb)^{\kb'-\kb},
\end{eqnarray}
and the probability $\br_{\kr, \kb}$ that an undeleted node reached by following a blue edge has $\kr$ red edges and $\kb+1$ blue edges after the attack as
\begin{eqnarray}
\label{def:abr}
\br_{\kr, \kb} 
=  \frac{1}{\tfb} \sum_{\kr'=\kr}^{\kcut; \df} p(\kr') \binom{\kr'}{\kr} \tfr^{\kr} (1-\tfr)^{\kr'-\kr} \sum_{\kb'=\kb}^{\infty} \tq(\kb') \binom{\kb'}{\kb} \tfb^{\kb} (1-\tfb)^{\kb'-\kb}.
\end{eqnarray}
The generating functions for $\bp_{\kr, \kb}$, $\bq_{\kr, \kb}$, and $\br_{\kr, \kb}$ are given from Eq.~(\ref{def:Fpqr}) with Eqs.~(\ref{def:abp})--(\ref{def:abr}):
\begin{subequations}
\begin{align}
F_p(x,y) &= \frac{1}{f} \sum_{\kr=\kmin}^{\kcut; \df} p(\kr) (\tfr x+1-\tfr)^{\kr} \tilde{G}_0 (\tilde{f}_{\rm B} y+1-\tilde{f}_{\rm B}), \label{def:aFp}  \\
F_q(x,y) &= \frac{1}{\tfr} \sum_{\kr=\kmin-1}^{\kcut-1; \df} q(\kr) (\tfr x+1-\tfr)^{\kr} \tilde{G}_0(\tilde{f}_{\rm B} y+1-\tilde{f}_{\rm B}), \label{def:aFq}  \\
F_r(x,y) &= \frac{1}{\tfb} \sum_{\kr=\kmin}^{\kcut; \df} p(\kr) (\tfr x+1-\tfr)^{\kr} \tilde{G}_1 (\tilde{f}_{\rm B} y+1-\tilde{f}_{\rm B}). \label{def:aFr}
\end{align}
\label{eq:F-uniform}
\end{subequations}
Using Eqs.~(\ref{def:bs_GC}) and (\ref{def:vrvb}) with Eqs.~(\ref{eq:F-uniform}), we evaluate $\bs_{\rm GC}$ for the attack on random networks with the uniform reinforcement.

Targeted attacks collapse the network at the critical threshold $f_c$: in the large size limit, $\bs_{\rm GC} > 0$ when $f>f_c$ and $\bs_{\rm GC}=0$ when $f \le f_c$.
We derive the critical threshold $f_c$ as follows.
Let us consider $\vred \approx 1-\ered$ and $\vb \approx 1-\eb$ to evaluate the stability of the trivial solution, $(\vred, \vb) = (1, 1)$.
Expanding Eqs.~(\ref{def:vrvb}) with Eqs.~(\ref{def:aFq}) and (\ref{def:aFr}) to the leading order in $\ered$ and $\eb$, we have
$(\ered \, \eb)^T = A \, (\ered \, \eb)^T$, where
\begin{equation}
A =
\begin{pmatrix}
   \sum_{\kr=\kmin-1}^{\kcut-1; \df} \kr q(\kr)  &  \quad \tfb \tilde{G}'_0(1)\\
   \frac{\tfr}{\tfb} \sum_{\kr=\kmin}^{\kcut; \df} \kr p(\kr)  & \quad \tfb \tilde{G}'_1(1)
\end{pmatrix}.
\label{def:aA}
\end{equation}
The trivial solution becomes unstable when the largest eigenvalue of $A$ exceeds one. 
Thus, the critical threshold $f_c$ is given by the following condition:
\begin{equation}
{\rm det}(A-I)=0,
\label{eq:det}
\end{equation} 
where $I$ is the identity matrix. 
Substituting Eq.~(\ref{def:aA}) into Eq.~(\ref{eq:det}) gives $f_c$ as the value of $f$ that satisfies 
\begin{equation}
(\tfb \tilde{G}'_1(1)-1 ) \left( \sum_{\kr=\kmin-1}^{\kcut-1; \df} \kr q(\kr) -1 \right) -\tfr \tilde{G}'_0(1) \left( \sum_{\kr=\kmin}^{\kcut; \df} \kr p(\kr) \right) = 0. \label{eq:fc-uniform}
\end{equation}

It should be noted that when $\tp(\kb)=\delta_{\kb,0}$ and $\tq(\kb)=0$, i.e., when a network is not reinforced, $\bs_{\rm GC}$ is given as follows:
\begin{equation}
\bs_{\rm GC} = f - \sum_{\kr=\kmin}^{\kcut; \df} p(\kr) (\tfr \vred+1-\tfr)^{\kr}, \label{eq:gc-original}
\end{equation}
where $\vred$ is the solution of
\begin{equation}
\vred = \frac{1}{\tfr} \sum_{\kr=\kmin-1}^{\kcut-1; \df} q(\kr) (\tfr \vred+1-\tfr)^{\kr}, \label{eq:v-original}
\end{equation}
and $f_c$ satisfies 
\begin{equation}
\sum_{\kr=\kmin-1}^{\kcut-1; \df} \kr q(\kr) -1 = 0. \label{eq:fc-original}
\end{equation}

\subsection{Selective Reinforcement} \label{sec:selective}

Next, we consider random networks with selective reinforcement where edges for reinforcement are added randomly to connect only minimum degree nodes.
Under selective reinforcement, nodes with $\kr=\kmin$ have $\kb$ blue edges with probability $\tp(\kb)$ and the other nodes have no blue edges.
Note that a node reachable by traversing a blue edge always has $\kr = \kmin$ under the selective reinforcement approach.
Then, the joint probabilities of the reinforced networks are expressed as follows:
\begin{subequations}
\begin{align}
p_{\kr, \kb} &= 
\begin{cases}
p(\kr) \delta_{\kb,0} & (\kr > \kmin) \\
p(\kr) \tp(\kb) & (\kr = \kmin)
\end{cases}, \\
q_{\kr, \kb} &= 
\begin{cases}
q(\kr) \delta_{\kb,0} & (\kr > \kmin-1) \\
q(\kr) \tp(\kb) & (\kr = \kmin-1)
\end{cases}, \\
r_{\kr, \kb} &= \delta_{\kr,\kmin} \tilde{q}(\kb) =
\begin{cases}
0 & (\kr > \kmin) \\
\tq(\kb) & (\kr = \kmin)
\end{cases},
\end{align}
\end{subequations}
where $\delta_{i,j}$ is the Kronecker delta.

Let us consider the targeted attack on selectively reinforced networks.
As for $f$ and $\tfr$, Eqs.~(\ref{def:f}) and (\ref{def:tfr}) hold in this case.
Noticing that nodes with $\kr = \kmin$ are never deleted as long as $f \ge p(\kmin)$, which is equivalent to $\kcut > \kmin$, we have for selective reinforcement,
\begin{equation}
\tfb =
\begin{cases}
1 & (f \geq p(\kmin)) \\
1-\Delta f & (f< p(\kmin))
\end{cases}.
\label{def:tfb}
\end{equation}

In the following, we describe the joint probabilities for undeleted nodes after the attack.
First, we consider $\bp_{\kr, \kb}$, i.e., the probability that an undeleted node has $\kr$ red edges and $\kb$ blue edges connected to undeleted nodes.
If $f > p(\kmin)$, it is sufficient to distinguish between the $\kr > \kmin$ and $\kr \leq \kmin$ cases.
All nodes with $\kr' \ge \kr > \kmin$ are not reinforced under selective reinforcement; thus, we obtain 
\begin{eqnarray} \label{def:bp11}
\bp_{\kr, \kb} &=& \frac{1}{f} \sum_{\kr'=\kr}^{\kcut; \df} p(\kr') \binom{\kr'}{\kr} \tfr^{\kr} (1-\tfr)^{\kr'-\kr}  \delta_{\kb,0} \quad (\kr > \kmin).
\end{eqnarray}
On the other hand, both unreinforced nodes and reinforced nodes can have $\kr \le \kmin$ red edges remaining after the attack; thus, we obtain
\begin{eqnarray} \label{def:bp12}
\bp_{\kr, \kb} &=& \frac{1}{f} \sum_{\kr'=\kmin+1}^{\kcut; \df} p(\kr') \binom{\kr'}{\kr} \tfr^{\kr} (1-\tfr)^{\kr'-\kr}  \delta_{\kb,0}  \nonumber \\
&& +\frac{1}{f} p(\kmin) \binom{\kmin}{\kr} \tfr^{\kr} (1-\tfr)^{\kmin-\kr} \sum_{\kb'=\kb}^\infty \tp(\kb') \binom{\kb'}{\kb} \tfb^{\kb} (1-\tfb)^{\kb'-\kb}  \quad (\kr \le \kmin).
\end{eqnarray}
When $f < p(\kmin)$, all unreinforced nodes are deleted due to the attack ($\kcut = \kmin$) and only a fraction $1-\df$ of reinforced nodes is retained.
Hence, $\bp_{\kr, \kb} = 0$ for $\kr > \kmin$ and 
\begin{eqnarray} \label{def:bp2}
\bp_{\kr,\kb} = \frac{1-\df}{f} p(\kmin) \binom{\kmin}{\kr} \tfr^{\kr} (1-\tfr)^{\kmin-\kr} \sum_{\kb'=\kb}^{\infty} \tp(\kb') \tfb^{\kb} (1-\tfb)^{\kb'-\kb}  
\end{eqnarray}
for $\kr \leq \kmin$.

Similarly, $\bq_{\kr, \kb}$ can be expressed as follows. 
When $f \geq p(\kmin)$, 
\begin{eqnarray} \label{def:bq11}
\bq_{\kr, \kb} = \frac{1}{\tfr} \sum_{\kr'=\kr}^{\kcut-1; \df} q(\kr') \binom{\kr'}{\kr} \tfr^{\kr} (1-\tfr)^{\kr'-\kr}  \delta_{\kb,0}
\end{eqnarray}
for $\kr > \kmin$ and 
\begin{eqnarray} \label{def:bq12}
\bq_{\kr, \kb}
&=& 
\frac{1}{\tfr} \sum_{\kr'=\kmin}^{\kcut-1; \df} q(\kr') \binom{\kr'}{\kr} \tfr^{\kr} (1-\tfr)^{\kr'-\kr} \delta_{\kb,0} \nonumber \\
&&
+\frac{1}{\tfr} q(\kmin-1) \binom{\kmin-1}{\kr} \tfr^{\kr} (1-\tfr)^{\kmin-1-\kr} \sum_{\kb'=\kb}^\infty \tp(\kb') \binom{\kb'}{\kb} \tfb^{\kb} (1-\tfb)^{\kb'-\kb}
\end{eqnarray}
for $\kr \leq \kmin$. 
When $f < p(\kmin)$, $\bq_{\kr,\kb} = 0$ for $\kr > \kmin$ and 
\begin{eqnarray} \label{def:bq21}
\bq_{\kr,\kb} = \frac{1-\df}{\tfr} q(\kmin-1) \binom{\kmin-1}{\kr} \tfr^{\kr} (1-\tfr)^{\kmin-1-\kr} \sum_{\kb'=\kb}^{\infty} \tp(\kb') \tfb^{\kb} (1-\tfb)^{\kb'-\kb} \qquad
\end{eqnarray}
for $\kr \leq \kmin$. 
In addition, $\br_{\kr,\kb}$ is as follows.
Since the nodes connected by blue edges always have the minimum degree in the original network, $\br_{\kr,\kb} = 0$ if $\kr > \kmin$.
For $\kr \le \kmin$, nodes with $\kr = \kmin$ in the original network are not subject to attacks and $\tfb=1$ as long as $f \ge p(\kmin)$.
Thus,
\begin{eqnarray} \label{def:br1}
\br_{\kr,\kb} &=& \binom{\kmin}{\kr} \tfr^{\kr}(1-\tfr)^{\kmin-\kr} \tilde{q}(\kb) \quad (\kr \le \kmin),
\end{eqnarray}
for $f \ge p(\kmin)$, whereas 
\begin{eqnarray} \label{def:br2}
\br_{\kr,\kb} &=& \binom{\kmin}{\kr} \tfr^{\kr}(1-\tfr)^{\kmin-\kr} \sum_{\kb'=\kb}^\infty \tilde{q}(\kb') \binom{\kb'}{\kb} \tfb^{\kb} (1-\tfb)^{\kb'-\kb} \quad (\kr \le \kmin),
\end{eqnarray}
for $f < p(\kmin)$.

The corresponding generating functions are summarized as follows.
When $f \geq p(\kmin)$,
\begin{subequations}
\begin{align}
F_p(x,y) &= \frac{1}{f} \left(\sum_{\kr'=\kmin+1}^{\kcut; \df} p(\kr') (\tfr x+1-\tfr)^{\kr'} + p(\kmin) (\tfr x+1-\tfr)^{\kmin} \tilde{G}_0 (\tfb y+1-\tfb) \right), \label{def:bFp1}\\
F_q(x,y) &=\frac{1}{\tfr} \left( \sum_{\kr'=\kmin}^{\kcut-1; \df} q(\kr') (\tfr x+1-\tfr)^{\kr'} + q(\kmin-1) (\tfr x+1-\tfr)^{\kmin-1} \tilde{G}_0 (\tfb y+1-\tfb) \right), \label{def:bFq1}\\
\intertext{and}
F_r(x,y) &= (\tfr x+1-\tfr)^{\kmin} \tilde{G}_1(y). \label{def:bFr1}
\end{align}
\end{subequations}
When $f<p(\kmin)$,
\begin{subequations}
\begin{align}
F_p(x,y) &= \frac{1-\df}{f} p(\kmin) (\tfr x+ 1-\tfr)^{\kmin} \tilde{G}_0(\tfb y+1-\tfb), \label{def:bFp2}\\
F_q(x,y) &= \frac{1-\df}{\tfr} q(\kmin-1) (\tfr x+ 1-\tfr)^{\kmin-1} \tilde{G}_0(\tfb y+1-\tfb), \label{def:bFq2}\\
\intertext{and}
F_r(x,y) &= (\tfr x+1-\tfr)^{\kmin} \tilde{G}_1(\tfb y+1-\tfb). \label{def:bFr2}
\end{align}
\end{subequations}
Substituting these generating functions into Eqs.~(\ref{def:bs_GC}) and~(\ref{def:vrvb}), we evaluate the normalized size $\bs_{\rm GC}$ of the GC for the attack on random networks with selective reinforcement.

The critical threshold $f_c$ of the targeted attack on selectively reinforced networks is evaluated according to the condition (\ref{eq:det}) with
\begin{equation} \label{eq:A1}
A =
\begin{pmatrix}
   \sum_{\kr=\kmin-1}^{\kcut-1; \df} \kr q(\kr) &  \quad  \frac{\tfr}{\tfb} q(\kmin-1)\tilde{G}'_0(1)\\
   \kmin \tfr & \quad \tilde{G}'_1(1)
\end{pmatrix}
\end{equation}
for $f \geq p(\kmin)$ and with
\begin{equation} \label{eq:A2}
A =
\begin{pmatrix}
  (\kmin-1)\tfr  &  \quad  \tfb \tilde{G}'_0(1)\\
   \kmin \tfr & \quad  \tfb \tilde{G}'_1(1)
\end{pmatrix}
\end{equation}
for $f < p(\kmin)$.
Substituting Eqs.~(\ref{eq:A1}) and (\ref{eq:A2}) into Eq.~(\ref{eq:det}), we obtain the critical threshold $f_c$ as the value of $f$ which satisfies
\begin{equation}
\left ( \sum_{\kr=\kmin-1}^{\kcut-1; \df} \kr q(\kr) - 1 \right) \left ( \tilde{G}'_1(1) -1\right) - \kmin \tfb q(\kmin-1)\tilde{G}'_0(1) = 0 \label{eq:fc-selective-1}
\end{equation}
if $f =f_c \geq p(\kmin)$ and
\begin{equation}
\left (( \kmin-1)\tfr -1 \right) \left ( \tfb \tilde{G}'_1(1) -1 \right) - \kmin \tfr \tfb \tilde{G}'_0(1) = 0 \label{eq:fc-selective-2}
\end{equation}
if $f =f_c < p(\kmin)$.

\section{Results} \label{sec:result}

\begin{figure}
\begin{center}
(a)
	\includegraphics[width=7.5cm]{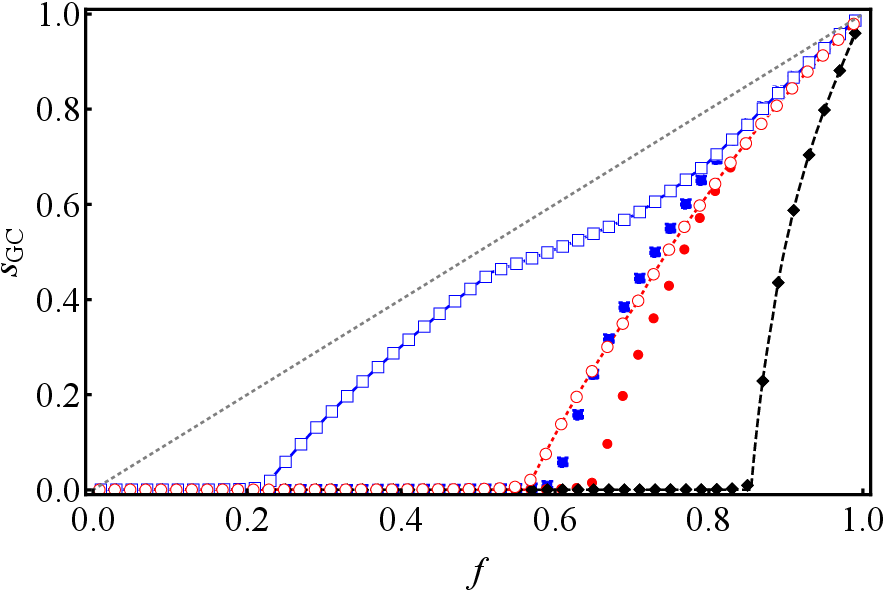}
(b)
	\includegraphics[width=7.5cm]{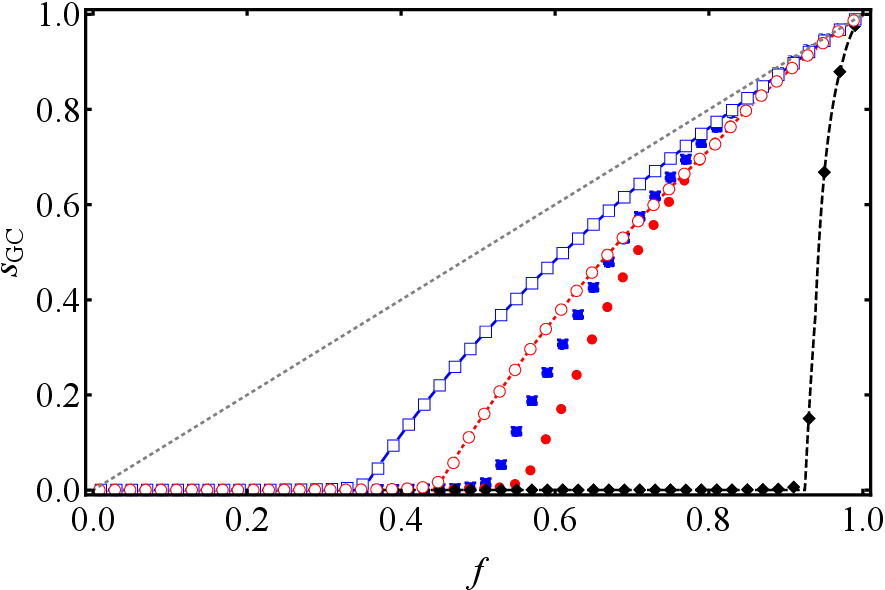}
\end{center}
\caption{
Normalized GC size $\bs_{\rm GC}$ of the targeted attacks on (un)reinforced scale-free networks with (a) $\gamma=2.5$ and (b) $\gamma=4.5$, as a function of the undeleted node fraction $f$.
The lines represent the analytical estimates based on the generating functions: 
the red dotted lines represent $\bs_{\rm GC}$ of the uniformly reinforced networks; 
the blue solid lines represent $\bs_{\rm GC}$ of the selectively reinforced networks;
and the black dashed lines represent $\bs_{\rm GC}$ of the unreinforced networks.
The symbols represent $\bs_{\rm GC}$ evaluated by Monte Carlo simulations with $N=10^5$ nodes:
the red open and filled circles represent the Monte Carlo results for the uniformly reinforced networks and their color-randomized counterparts, respectively;
the blue open and filled squares represent the Monte Carlo results for the selectively reinforced networks and their color-randomized counterparts, respectively;
and the black diamonds represent the Monte Carlo results for the unreinforced networks.
}
\label{fig:fbs_GC}
\end{figure}

\begin{table}
\begin{center}
\begin{tabular}{| l || c c c c | c c c c |} \hline
	 & \multicolumn{4}{c|}{$\gamma=2.5$} & \multicolumn{4}{c|}{$\gamma=4.5$} \\ 
	 &  $f_c$  &  $R$  & $f_c^{\rm (rand)}$  &  $R^{\rm (rand)}$  & $f_c$  &  $R$  & $f_c^{\rm (rand)}$  &  $R^{\rm (rand)}$  \\ \hline
	\, No reinforcement \,&\, 0.855 \,&\, 0.092 \,&\, 0.037 \,&\, 0.412 \,&\, 0.925 \,&\, 0.053 \,&\, 0.590 \,&\, 0.213 \,\\ 
	\, Uniform reinforcement \,&\, 0.563 \,&\, 0.243 \,&\, 0.037 \,&\, 0.440 \,&\, 0.445 \,&\, 0.313 \,&\, 0.344 \,&\, 0.373 \,\\ 
	\, Selective reinforcement \,&\, 0.220 \,&\, 0.400 \,&\, 0.037 \,&\, 0.444 \,&\, 0.350 \,&\, 0.367 \,&\, 0.357 \,&\, 0.372 \,\\ \hline
\end{tabular}
\end{center}
\caption{Critical threshold $f_c$ and robustness measure $R$ of the targeted attacks and random failures on scale-free networks in each setup.
The (red) degree distribution of scale-free networks is $p(\kr) = \kr^{-\gamma} / \sum_{k'=2}^{500} k'^{-\gamma}$ ($2 \le \kr \le 500$).
The values $f_c$ and $R$ represent the critical threshold and the robustness measure for the targeted attacks, respectively, which were evaluated using the equations derived in Section~\ref{sec:analysis}; $f_c^{\rm (rand)}$ and $R^{\rm (rand)}$ represent the critical threshold and the robustness measure for the random failures, respectively, which were evaluated using the equations derived in Appendix~\ref{sec:randomFailure}.
}
\label{tab:fcR}
\end{table}

In this section, we discuss the extent to which the uniform and selective reinforcement approaches improve the robustness of scale-free networks against targeted attacks.
In this evaluation, we prepare uncorrelated scale-free networks with degree distribution 
\be
p(\kr) = c \kr^{-\gamma} \quad (\kr=\kmin, \kmin+1, \cdots, \kmax) \label{eq:degdis}
\ee
as the original networks.
Here $c$ is the normalizing constant, i.e., $c=1/\sum_{\kr = \kmin}^{\kmax} \kr^{-\gamma}$.
We set the minimum degree to $\kmin =2$ and the maximum degree to $\kmax = 500$.  
Under uniform reinforcement, edges are added to all nodes at random. 
The blue degree distribution is then $\tp(\kb) = z_{\rm rand}^{\kb} e^{-z_{\rm rand}}/\kb!$. 
In this study, we set $z_{\rm rand}=1$ so that the number of blue edges per node is 1.
Under selective reinforcement, only the minimum degree nodes have blue degree $\kb$ with probability $\tp(\kb) = z_{\rm sele}^{\kb} e^{-z_{\rm sele}}/\kb!$. 
We ensure the same number of blue edges in both uniform and selective reinforcement by setting $z_{\rm sele} = z_{\rm rand} / p(k_{\rm min})$. 
In a uniformly reinforced network with $N$ nodes, blue edges are randomly added to all nodes, resulting in a total number of blue edges of $N z_{\rm rand}/2$, while in a selectively reinforced network with $N $ nodes, blue edges are randomly added to the minimum degree nodes, resulting in a total number of blue edges of $N p(k_{\rm min}) z_{\rm sele}/2$.

We ran Monte Carlo simulations along with the analytical estimates derived in the previous section. 
In the simulations, we generated $10^4$ network realizations, each with $N=10^5$ nodes.
For each realization, we implemented the Newman--Ziff algorithm~\cite{newman2000efficient,newman2001fast} for the targeted attack to evaluate the fraction of the largest component over the entire range from $f=0$ to $f=1$.
The fraction of the largest component averaged over all runs was taken as $\bs_{\rm GC}$.
In our simulations, we also employed the color-randomized counterparts for the uniformly reinforced and selectively reinforced networks. 
After generating a reinforced network, we reassigned the color (red or blue) of each edge while keeping the total number of red and blue edges unchanged, resulting in a color-randomized network.

Figures~\ref{fig:fbs_GC}(a) and~\ref{fig:fbs_GC}(b) plot the normalized GC size $\bs_{\rm GC}$ of the targeted attacks on (un)reinforced scale-free networks with $\gamma=2.5$ and $\gamma=4.5$, respectively.
Here the average red degree is $\langle \kr \rangle \approx 4.46$ when $\gamma=2.5$ and $\langle \kr \rangle \approx 2.32$ when $\gamma=4.5$.
For both $\gamma=2.5$ and $\gamma=4.5$, the scale-free networks without reinforcement are vulnerable to the targeted attacks.
In these figures, the theoretical lines based on the generating functions and the Monte Carlo results (symbols) are in perfect agreement for both the uniform and selective reinforcement approaches, confirming the validity of our formulation.

We observe from Figs.~\ref{fig:fbs_GC}(a) and (b) that the reinforced networks, either uniform or selective, become more robust against attacks than the unreinforced networks.
This seems obvious because the reinforced networks have more edges than the unreinforced networks and their degree distribution is altered by the reinforcement, so that the average degree increases.
We therefore compared the uniformly reinforced networks and selectively reinforced networks with their color-randomized counterparts.
The comparisons clearly show that both reinforced networks exhibit more robustness than their color-randomized counterparts.

Table~\ref{tab:fcR} shows the critical threshold and the robustness measure of the targeted attacks and random failures, where an $f$ fraction of randomly selected nodes are retained and the other nodes are deleted, on unreinforced and reinforced networks.
Here the robustness measure~\cite{schneider2011mitigation} is defined as
\begin{equation}
R=\frac{1}{N}\sum_{Q=1}^{N}\bs_{\rm GC}(Q),
\end{equation}
where $\bs_{\rm GC}(Q)$ is the fraction of the largest component after removing $Q=(1-f)N$ nodes.
Here, the range of $R$ is $1/N \le R \le 1/2$ by definition.
Both $R$ and $f_c$ quantify the robustness of a network against targeted attacks: for small $f_c$ values, the GC exists even when many nodes are removed, which means that the network is robust; for large $R$ values, the GC having a large relative size is preserved against node removals, which means that the network is robust.
The $f_c$ and $R$ values in the table show that the selective reinforcement improves the robustness of the scale-free networks against targeted attacks significantly.
Furthermore, the selective reinforcement appears to be more effective in more heterogeneous networks, as suggested by the comparison between the $\gamma = 2.5$ and $\gamma = 4.5$ cases.
As a side note, there is little difference between the selective and uniform reinforcement approaches in terms of improving the robustness against random failures (see also Appendix~\ref{sec:randomFailure}).

\begin{figure}
\begin{center}
(a)
	\includegraphics[width=7.5cm]{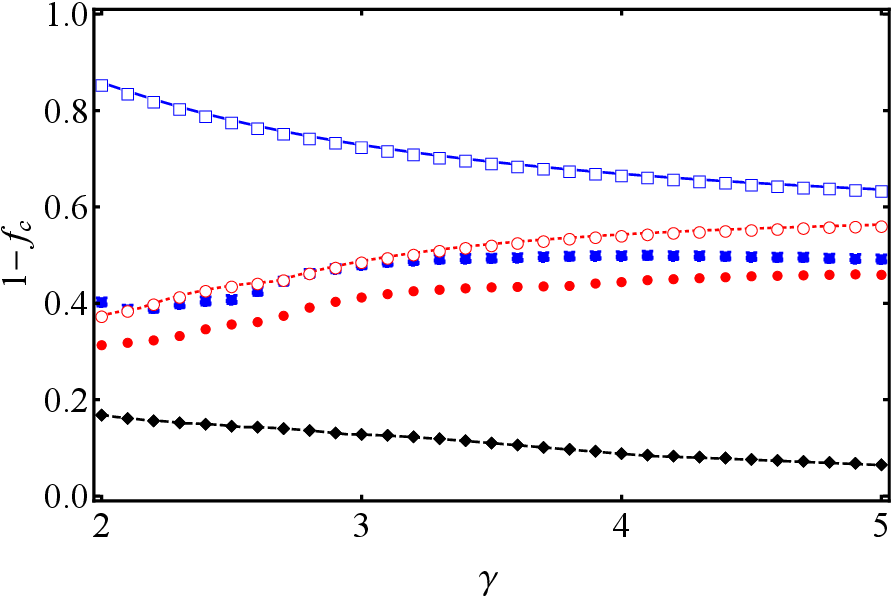}
(b)
	\includegraphics[width=7.5cm]{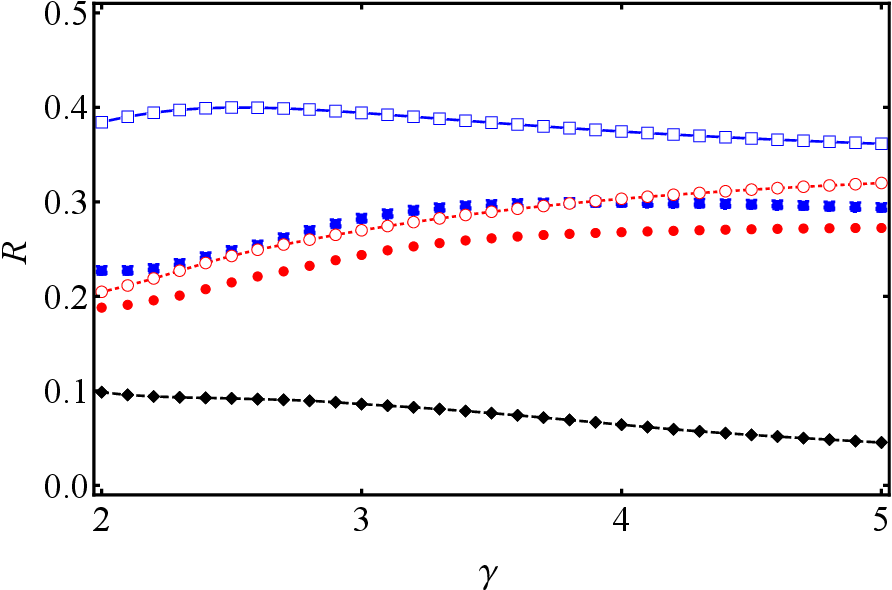}
\end{center}
\caption{
(a) Critical threshold $1-f_c$ and (b) robustness measure $R$ of the targeted attacks on (un)reinforced scale-free networks, as a function of the degree exponent $\gamma$.
The original networks have $\kmin=2$ and $\kmax=500$.
The red dotted lines, blue solid lines, and black dashed lines represent the analytical estimates based on the generating functions for the uniformly reinforced networks, selectively reinforced networks, and unreinforced networks, respectively. 
The symbols represent $1-f_c$ and $R$ evaluated by Monte Carlo simulations of $10^3$ networks with $N=10^6$ nodes:
the red open and filled circles represent the Monte Carlo results for the uniformly reinforced networks and their color-randomized counterparts, respectively;
the blue open and filled squares represent the Monte Carlo results for the selectively reinforced networks and their color-randomized counterparts, respectively;
the black diamonds represent the Monte Carlo results for the unreinforced networks.
In these simulations, $f_c$ was evaluated as the value of $f$, at which $\bs_{\rm GC}=0.01$.
}
\label{fig:gammadependence}
\end{figure}

Finally, we discuss how the performance of reinforcement is affected when the network heterogeneity varies.
Figures~\ref{fig:gammadependence}(a) and~\ref{fig:gammadependence}(b) show the critical threshold $1-f_c$ and the robustness measure $R$ of the (un)reinforced networks and their color-randomized counterparts, as a function of the degree exponent $\gamma$.
Both in terms of $R$ and $1-f_c$, the uniformly reinforced and selectively reinforced networks are more robust (i.e., they have larger $1-f_c$ and larger $R$ values) than their color-randomized counterparts, regardless of $\gamma$.
As the figures indicate, the selectively reinforced networks are the most robust, and their superiority is more pronounced when $\gamma$ is smaller, i.e., when the given networks are more heterogeneous.
In particular, it should be noted that only the selectively reinforced networks increase $1-f_c$ as $\gamma$ decreases.
This is because as $\gamma$ decreases in a pure power-law degree distribution, the fraction of the minimum degree nodes is reduced and the average number of blue edges per minimum degree node is increased, thereby resulting in a tightly connected subgraph of minimum degree nodes.

We performed similar calculations by changing $k_{\rm min}$ in the power-law degree distribution (\ref{eq:degdis}) to $1, 3$ and $4$. 
Our results remain qualitatively unchanged for different values of $k_{\rm min}$ (not shown).
All of the presented results reveal that the selective reinforcement approach is effective in improving the robustness of scale-free networks against targeted attacks.


\begin{figure}
\begin{center}
(a)
	\includegraphics[width=7.5cm]{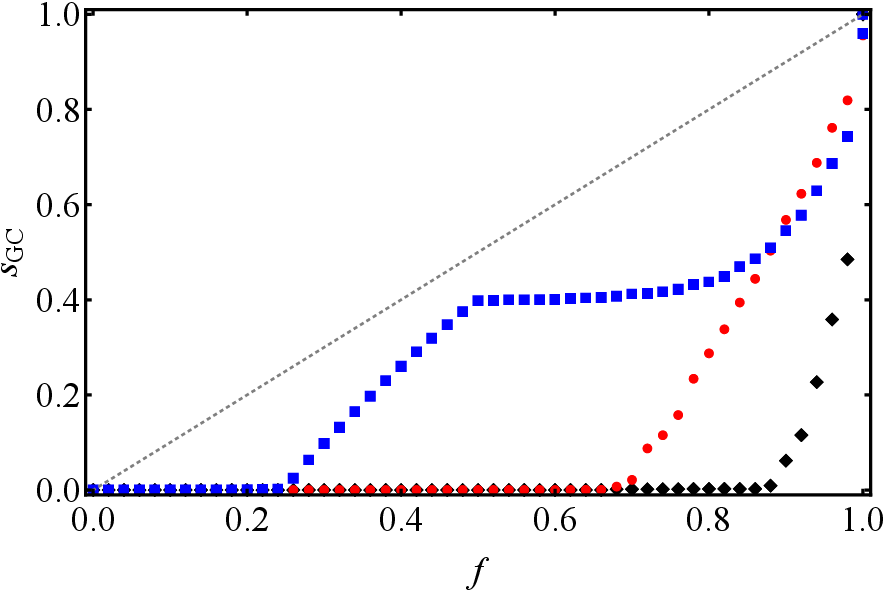}
(b)
	\includegraphics[width=7.5cm]{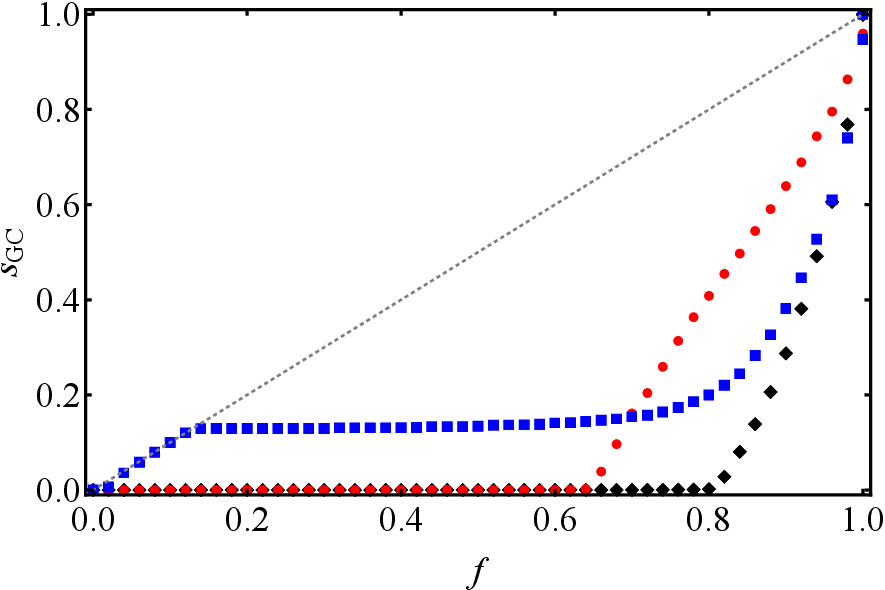}
\end{center}
\caption{
Effect of reinforcement on the normalized GC size $\bs_{\rm GC}$ of the targeted attacks on real networks: (a) Notre Dame web graph~\cite{albert1999diameter,snapnets}, where the number of nodes is $N=325729$ and the average (red) degree is $\langle \kr \rangle \approx 7$; (b) Internet graph~\cite{leskovec2005graphs,snapnets}, where the number of nodes is $N=1696415$ and the average (red) degree is $\langle \kr \rangle \approx 14$.
The fraction of minimum degree nodes is $p(\kmin=1)\approx 0.50$ for the web graph and $p(\kmin=1)\approx 0.13$ for the Internet graph.
For simplicity, the former network was converted to an undirected network.
The black diamonds, the red circles and the blue squares represent the results without reinforcement, with uniform reinforcement, and with selective reinforcement, respectively.
}
\label{fig:realnetwork}
\end{figure}

\section{Summary and Discussion} \label{sec:summary}

In this paper, we investigated the robustness of scale-free networks against targeted attacks when the networks are reinforced by adding hidden edges.
We formulated the normalized size of the GC and the critical threshold for the targeted attacks on random networks with uniform reinforcement, where all nodes in a network are targeted for edge additions, and with selective reinforcement, where the minimum degree nodes are targeted for edge additions, using generating functions.
Applying our analysis and Monte Carlo simulations to the targeted attacks on scale-free networks, we found that both the uniform and selective reinforcement approaches render scale-free networks robust.
In particular, selective reinforcement significantly improves network robustness in terms of both the critical threshold and the robustness measure. 
The results of analytical estimates obtained using the generating functions and Monte Carlo simulations are in perfect agreement for both the uniform and selective reinforcement approaches, confirming the validity of our formulation.

We should mention here the degree correlation~\cite{newman2002assortative,newman2003mixing} of the reinforced networks, in that the relation between the degree correlation and the network robustness against targeted attacks has been discussed in the literature~\cite{shiraki2010cavity,schneider2011mitigation,herrmann2011onion,tanizawa2012robustness,mizutaka2016robustness,wu2011onion}.
In the present case, where the given networks are degree-uncorrelated, the assortativity coefficient of the reinforced networks can be formulated using the generating functions (Appendix~\ref{sec:assortativity}).
As shown in Fig.~\ref{fig:assortativity} (Appendix~\ref{sec:assortativity}), the degree correlation of the reinforced networks is positive, but very weakly positive (moreover, the assortativity coefficient of the reinforced networks approaches zero as $\kmax$ increases for $\gamma \le 4$). 
Both the uniform and selective reinforcement approaches increase the number of connections between low degree nodes; however, this increase is insufficient to change the degree correlation of the given network significantly.
The question of whether the selective reinforcement is effective for correlated networks was not addressed in this study. 
This will be investigated and discussed in a future study.

Finally, we present preliminary results that confirm the effectiveness of selective reinforcement for real networks.
Figures~\ref{fig:realnetwork}(a) and~\ref{fig:realnetwork}(b) plot the largest component fraction $\bs_{\rm GC}$ in the targeted attacks on a web graph and an Internet graph, respectively, as a function of the undeleted node fraction $f$.
In both networks, it is observed that selective reinforcement significantly reduces the critical threshold, above which $\bs_{\rm GC}$ is non-negligibly positive.
The robustness measure is also higher for selective reinforcement than for both no reinforcement and uniform reinforcement.
However, we should be aware of the outcome of the selectively reinforced Internet graph (the blue squares in Fig.~\ref{fig:realnetwork}(b)).
In the selectively reinforced Internet graph, $\bs_{\rm GC}$ drops as sharply as in the unreinforced Internet graph due to the removal of a few high degree nodes. 
It then falls well below the uniformly reinforced case.
On the other hand, after $\bs_{\rm GC}$ reaches approximately 0.13, which corresponds to the fraction of the minimum degree nodes, it remains flat, and the GC exists until nearly of the nodes have been removed.
For cases where the number of minimum degree nodes is small, the current selective reinforcement may be redundant in the sense that it adds an excessive number of hidden edges to the limited number of nodes.
For real networks, which generally do not follow a pure power-law degree distribution, selective reinforcement should be tuned to maximize the robustness parameter.
In this argument, we did not consider practical issues such as the cost and feasibility of adding blue edges. 
Constraints such as the number of allowable blue edges and their lengths will limit the application of our reinforcement approaches to real networks, especially when a network is embedded in geographical space. 
The very idea of hidden edges may also be difficult to implement in some situations. 
Therefore, while our study provides useful guidelines for improving network robustness, further research is needed to address these practical considerations and optimize selective reinforcement for real networks.

\section*{Acknowledgment}

We thank Toshihiro Tanizawa and Shogo Mizutaka for their useful comments and suggestions.
This work was supported by JSPS KAKENHI Grant Numbers JP19K03648 and JP21H03425.

 
 \appendix

\section{Robustness of Reinforced Networks against Random Failures} \label{sec:randomFailure}

In this appendix, we derive the normalized GC size and the critical threshold for random failures on reinforced networks.
In random failures on a reinforced network, each node is retained with probability $f$ and otherwise deleted, regardless of its red degree and blue degree.
Thus, for random failures, 
\be
f = \tfr = \tfb.
\ee
The results for the uniform and selective reinforcement cases are described below.

\paragraph{Uniform reinforcement}
The joint probabilities for undeleted nodes after a random failure are expressed as follows:
\begin{subequations}
\begin{align}
\bp_{\kr, \kb} &= \sum_{\kr'=\kr}^{\infty} p(\kr') \binom{\kr'}{\kr} f^{\kr} (1-f)^{\kr'-\kr} \sum_{\kb'=\kb}^{\infty} \tp(\kb') \binom{\kb'}{\kb} f^{\kb} (1-f)^{\kb'-\kb},
\label{def:randomabp}
\\
\bq_{\kr, \kb} &= \sum_{\kr'=\kr}^{\infty} q(\kr') \binom{\kr'}{\kr} f^{\kr} (1-\tf)^{\kr'-\kr} \sum_{\kb'=\kb}^{\infty} \tp(\kb') \binom{\kb'}{\kb} f^{\kb} (1-f)^{\kb'-\kb},
\label{def:randomabq}
\\
\intertext{and}
\br_{\kr, \kb} &= \sum_{\kr'=\kr}^{\infty} p(\kr') \binom{\kr'}{\kr} f^{\kr} (1-f)^{\kr'-\kr} \sum_{\kb'=\kb}^{\infty} \tq(\kb') \binom{\kb'}{\kb} f^{\kb} (1-f)^{\kb'-\kb}.
\label{def:randomabr}
\end{align}
\end{subequations}
The corresponding generating functions are
\begin{subequations}
\begin{align}
F_p(x,y) &= G_0 (fx+1-f) \tilde{G}_0 (f y+1-f), \\
F_q(x,y) &= G_1 (fx+1-f) \tilde{G}_0 (f y+1-f), \\
F_r(x,y) &= G_0 (fx+1-f) \tilde{G}_1 (f y+1-f).
\end{align}
\label{eq:randomF}
\end{subequations}
Substituting Eqs.~(\ref{eq:randomF}) into Eqs.~(\ref{def:bs_GC}) and~(\ref{def:vrvb}) allows us to evaluate $\bs_{\rm GC}$.
The critical threshold $f_c$ is determined according to condition (\ref{eq:det}) with matrix 
\be
 A =
 \begin{pmatrix}
   f G'_1(1) & \quad f \tilde{G}'_0(1)\\
   f G'_0(1) & \quad f \tilde{G}'_1(1)
\end{pmatrix},
\label{def:AppendixaA}
\ee
from which $f_c$ is the value of $f$ that satisfies
\begin{equation}
 (f G'_1(1)-1 ) (f \tilde{G}'_1(1)-1 ) - f^2 G'_0(1) \tilde{G}'_0(1) = 0.
\end{equation}

\paragraph{Selective reinforcement}
The joint probabilities for undeleted nodes after a random failure are expressed as follows.
First, $\bp_{\kr, \kb}$ is
\be
\bp_{\kr, \kb} = \sum_{\kr'=\kr}^{\infty} p(\kr') \binom{\kr'}{\kr} f^{\kr} (1-f)^{\kr'-\kr} \delta_{\kb,0}
\ee
for $\kr > \kmin$ and
\ba
\bp_{\kr, \kb}
&=& 
\sum_{\kr'=\kmin+1}^{\infty} p(\kr') \binom{\kr'}{\kr} f^{\kr} (1-f)^{\kr'-\kr} \delta_{\kb,0}
 \nonumber \\
&&
+ p(\kmin) \binom{\kmin}{\kr} f^{\kr} (1-f)^{\kmin-\kr} \sum_{\kb'=\kb}^\infty \tp(\kb') \binom{\kb'}{\kb} \tfb^{\kb} (1-\tfb)^{\kb'-\kb}  
\ea
for $\kr \leq \kmin$.
Similarly, $\bq_{\kr, \kb}$ is
\be
\bq_{\kr, \kb} = \sum_{\kr'=\kr}^{\infty} q(\kr') \binom{\kr'}{\kr} f^{\kr} (1-f)^{\kr'-\kr} \delta_{\kb,0}
\ee
for $\kr > \kmin - 1$ and
\ba
\bq_{\kr, \kb}
&=& 
\sum_{\kr'=\kmin}^{\infty} q(\kr') \binom{\kr'}{\kr} f^{\kr} (1-f)^{\kr'-\kr} \delta_{\kb,0}
 \nonumber \\
&&
+ q(\kmin-1) \binom{\kmin-1}{\kr} f^{\kr} (1-f)^{\kmin-1-\kr} \sum_{\kb'=\kb}^\infty \tp(\kb') \binom{\kb'}{\kb} f^{\kb} (1-f)^{\kb'-\kb}  
\ea
for $\kr \leq \kmin - 1$.
Finally, $\br_{\kr,\kb} = 0$ for $\kr > \kmin$ and  
\be
\br_{\kr,\kb}= \binom{\kmin}{\kr} f^{\kr}(1-f)^{\kmin-\kr} 
\sum_{\kb'=\kb}^\infty \tilde{q}(\kb') \binom{\kb'}{\kb} f^{\kb} (1-f)^{\kb'-\kb}
\ee
for $\kr \leq \kmin$.

The corresponding generating functions are 
\begin{subequations}
\begin{align}
F_p(x,y) &= G_0 (fx+1-f) + p(\kmin) (fx+1-f)^{\kmin} (\tilde{G}_0 (f y+1-f)-1), \\
F_q(x,y) &= G_1 (fx+1-f) + q(\kmin-1) (fx+1-f)^{\kmin-1} (\tilde{G}_0 (f y+1-f)-1), \\
F_r(x,y) &= (fx+1-f)^{\kmin} \tilde{G}_1 (f y+1-f).
\end{align} \label{eq:selectiveF}
\end{subequations}
Substituting Eqs.~(\ref{eq:selectiveF}) into Eqs.~(\ref{def:bs_GC}) and~(\ref{def:vrvb}) allows us to evaluate $\bs_{\rm GC}$.
The critical threshold $f_c$ is determined according to condition (\ref{eq:det}) with matrix
\be
A =
\begin{pmatrix}
   f G'_1(1)  &  \quad  f q(\kmin-1)\tilde{G}'_0(1)\\
   f \kmin& \quad f \tilde{G}'_1(1)
\end{pmatrix},
\ee
from which $f_c$ is the value of $f$ that satisfies
\be
(f G'_1(1)-1 ) (f \tilde{G}'_1(1)-1 ) - f^2 \kmin q(\kmin-1) \tilde{G}'_0(1) = 0.
\ee

\begin{figure}
\begin{center}
(a)
	\includegraphics[width=7.5cm]{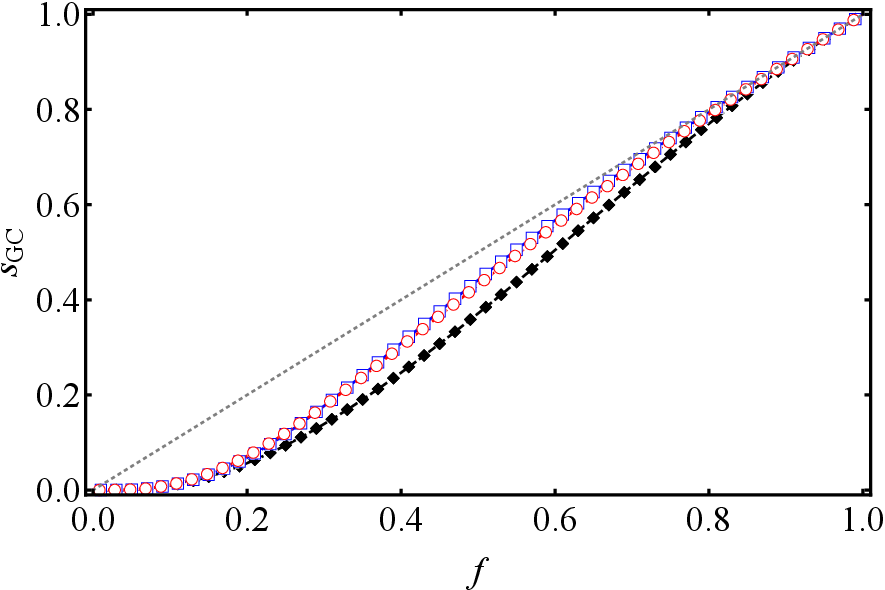}
(b)
	\includegraphics[width=7.5cm]{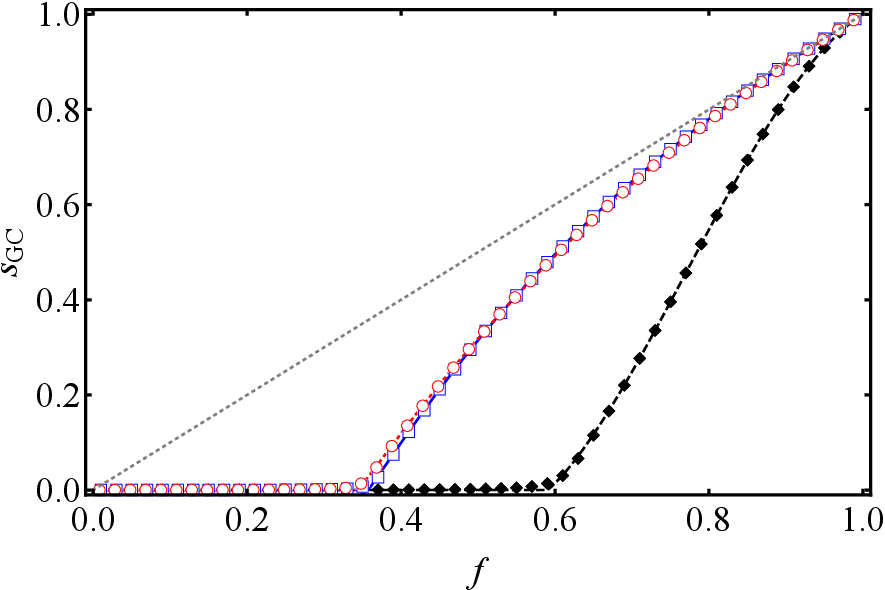}
\end{center}
\caption{
Normalized GC size $\bs_{\rm GC}$ of random failures on (un)reinforced scale-free networks with (a) $\gamma=2.5$ and (b) $\gamma=4.5$, as a function of the undeleted node fraction $f$.
The lines represent the analytical estimates based on the generating functions: 
the black dashed lines, the red dotted lines, and the blue solid lines represent $\bs_{\rm GC}$ of the unreinforced networks, the uniformly reinforced networks, and the selectively reinforced networks, respectively.
The symbols represent $\bs_{\rm GC}$ evaluated by Monte Carlo simulations with $N=10^5$ nodes:
the black diamonds, the red open circles, and the blue open squares represent the Monte Carlo results for the unreinforced networks, the uniformly reinforced networks, and the selectively reinforced networks, respectively.
In the simulations, the Newman--Ziff algorithm was employed for $10^4$ networks to evaluate the largest component fraction over the entire range of $f$.
The largest component fraction averaged over all runs was taken as $\bs_{\rm GC}$.
}
\label{fig:randomfailure}
\end{figure}

Figures~\ref{fig:randomfailure}(a) and~\ref{fig:randomfailure}(b) show the normalized GC size $\bs_{\rm GC}$ of random failures on (un)reinforced scale-free networks with $\gamma=2.5$ and $\gamma=4.5$, respectively.
As can be seen, the theoretical lines are in good agreement with the results of Monte Carlo simulations, which confirms the validity of the formulation presented here.

For $\gamma=2.5$, where the original network is very robust ($f_c \to 0$ as $N \to \infty$ by setting $\kmax$ unbounded) to random failures~\cite{callaway2000network,cohen2000resilience}, the reinforced one is by definition very robust (Fig.~\ref{fig:randomfailure}(a)).
The original network is so robust that it does not require reinforcement; thus, the increase in $\bs_{\rm GC}$ due to reinforcement is very small regardless of whether it is uniform or selective. 
The robustness measure $R$ increases from $R \approx 0.4120$ (no reinforcement) to $R \approx 0.4405$ (uniform reinforcement) or $R \approx 0.4441$ (selective reinforcement).

For $\gamma=4.5$, the edge additions improve the robustness of the network, both for the uniform and selective reinforcement cases (Fig.~\ref{fig:randomfailure}(b)): the critical threshold decreases from $f_c \approx 0.5903$ (no reinforcement) to $f_c \approx 0.3439$ (uniform reinforcement) or $f_c \approx 0.3572$ (selective reinforcement); and the robustness measure increases from $R \approx 0.2134$ (no reinforcement) to $R \approx 0.3726$ (uniform reinforcement) or $R \approx 0.3717$ (selective reinforcement). 
For a scale-free network, its high robustness to random failures is attributed to high degree nodes; however, these nodes are never or rarely selected as targets for edge additions, under the uniform and selective reinforcement approaches.
Thus, the difference in the improvement in the robustness between the uniform and selective reinforcement approaches is nearly negligible.

\section{Assortativity of Reinforced Networks} \label{sec:assortativity}

In this appendix, we formulate the assortativity coefficient of random networks with uniform reinforcement and selective reinforcement and discuss the assortativity of reinforced scale-free networks.
By definition, the assortativity coefficient of a network is given by the probability $Q(k, k')$ that two ends of a randomly selected edge have degree $k+1$ and $k'+1$ and the probability $Q(k)=\sum_{k'} Q(k,k')$ that an edge is connected to a node of degree $k+1$.
By introducing the generating function $B(x,y) = \sum_{k} \sum_{k'} Q(k,k') x^{k} y^{k'}$ for $Q(k,k')$ and $S(x) = B(x,1) = \sum_{k}Q(k)x^k$ for $Q(k)$, the assortativity coefficient $r_0$ is expressed as follows:
\begin{equation} \label{eq:r0}
r_0 = \frac{\sum_{k}\sum_{k'} k k' Q(k,k') -\left(\sum_k k Q(k)\right)^2}{\sum_{k} k^2 Q(k) -\left(\sum_k k Q(k)\right)^2}
= \frac{\partial_x \partial_y B(x,y) -(\partial_x S(x))^2}{(x \partial_x)^2 S(x) -(\partial_x S(x))^2}\Big|_{x=y=1}.
\end{equation}

For reinforced networks, there are two types of edges, i.e., red and blue edges.
We denote the number of red edges and the number of blue edges by $M_{\rm R}$ and $M_{\rm B}$ respectively.
The probability $\pr$ ($\pb$) that a randomly selected edge is red (blue) is
\begin{equation}
\pr = \frac{M_{\rm R}}{M_{\rm R} + M_{\rm B}} \quad \left( \pb = 1-\pr = \frac{M_{\rm B}}{M_{\rm R} + M_{\rm B}} \right).
\end{equation}
Let $Q_2(\kr, \kb,\kr', \kb')$ denote the probability that one end of a randomly selected edge has $\kr$ red edges and $\kb$ blue edges other than the selected edge, and the other end has $\kr'$ red edges and $\kb'$ blue edges other than the selected edge.
This probability is given as $Q_2(\kr, \kb,\kr', \kb') = \qkk \qkkp$ if the edge is red one and $Q_2(\kr, \kb,\kr', \kb') = \rkk \rkkp$ if the edge is blue one.
Introducing the generating function $B_2(\xr,\xb,\yr,\yb)$ for $Q_2(\kr, \kb,\kr', \kb')$, we have
\begin{eqnarray}
B_2(\xr,\xb,\yr,\yb) 
&=& \sum_{\kr}\sum_{\kb}\sum_{\kr'}\sum_{\kb'} Q_2(\kr \kb,\kr' \kb') \xr^{\kr} \xb^{\kb} \yr^{\kr'} \yb^{\kb'} \nonumber \\
&=& \sum_{\kr} \sum_{\kb} \sum_{\kr'} \sum_{\kb'} \left( \pr \qkk \qkkp + \pb \rkk \rkkp \right) \xr^{\kr} \xb^{\kb} \yr^{\kr'} \yb^{\kb'}.
\end{eqnarray}
We also introduce the probability $Q_2(\kr, \kb)$ that one end of an edge has $\kr$ red edges and $\kb$ blue edges other than the one and the corresponding generating function as 
\begin{eqnarray}
S_2(\xr,\xb) 
&=& \sum_{\kr}\sum_{\kb} Q_2(\kr, \kb) \xr^{\kr} \xb^{\kb} \nonumber \\
&=& B_2(\xr,\xb,1,1) \nonumber \\
&=& \sum_{\kr} \sum_{\kb} (\pr \qkk + \pb \rkk) \xr^{\kr} \xb^{\kb}.
\end{eqnarray}

Noticing that the degree of a node is the sum of its red degree and blue degree, we have
\begin{equation} \label{def:Bxy}
B(x,y) = B_2(x,x,y,y) \quad {\rm and} \quad S(x) = S_2(x,x).
\end{equation}
In the following, we describe $B(x,y)$ and $S(x)$ in the uniform and selective reinforcement cases.

\paragraph{Uniform reinforcement}
In the uniform reinforcement case, $B(x,y)$ is 
\begin{eqnarray} \label{def:aBxy}
B(x,y)
&=& \sum_{\kr} \sum_{\kb} \sum_{\kr'} \sum_{\kb'} \left(\pr q(\kr)\tp(\kb) q(\kr')\tp(\kb') + \pb p(\kr)\tq(\kb) p(\kr')\tq(\kb') \right) x^{\kr+\kb}  y^{\kr'+\kb'}\nonumber \\
&=& \pr G_1(x) \tilde{G}_0(x) G_1(y) \tilde{G}_0(y) + \pb G_0(x) \tilde{G}_1(x) G_0(y) \tilde{G}_1(y),
\end{eqnarray}
and $S(x)$ is 
\begin{eqnarray}
\label{def:aSx}
S(x) = \pr G_1(x) \tilde{G}_0(x) + \pb G_0(x) \tilde{G}_1(x).
\end{eqnarray}

\paragraph{Selective reinforcement}
In the selective reinforcement case, $B(x,y)$ is
\begin{eqnarray}
\label{def:bBxy}
B(x,y)
&=&
\pr \left (G_1(x) + q(\kmin-1) x^{\kmin-1}(\tilde{G}_0(x) - 1) \right)  \left (G_1(y) + q(\kmin-1) y^{\kmin-1}(\tilde{G}_0(y) - 1) \right) \nonumber \\
&& + \pb x^{\kmin} \tilde{G}_1(x) y^{\kmin} \tilde{G}_1(y),
\end{eqnarray}
and $S(x)$ is
\begin{eqnarray} \label{def:bSx}
S(x) = \pr \left (G_1(x) + q(\kmin-1) x^{\kmin-1}(\tilde{G}_0(x) - 1) \right) + \pb x^{\kmin} \tilde{G}_1(x).
\end{eqnarray}

\begin{figure}
\begin{center}
(a)
	\includegraphics[width=7.5cm]{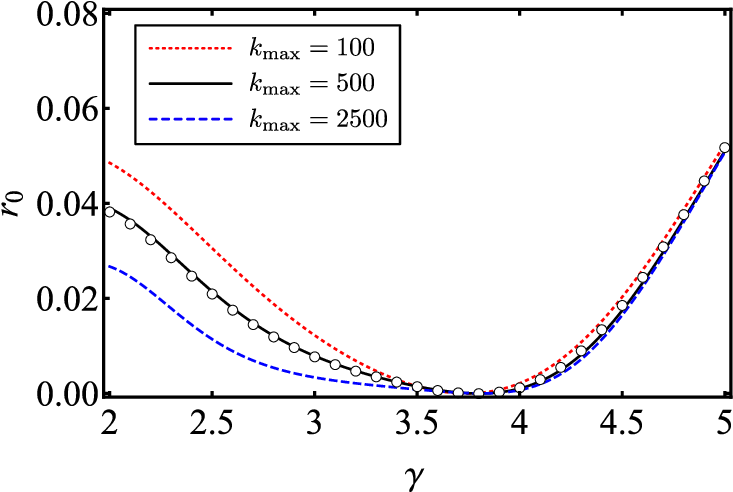}
(b)
	\includegraphics[width=7.5cm]{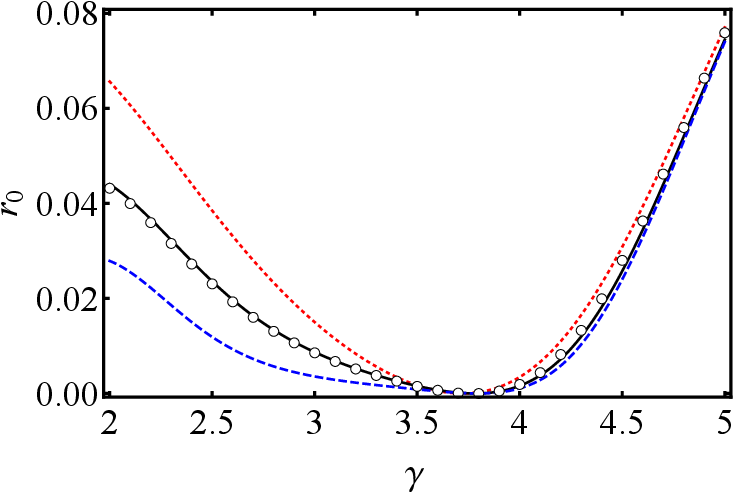}
\end{center}
\caption{
Assortativity coefficient $r_0$ of (a) uniformly reinforced scale-free networks and (b) selectively reinforced scale-free networks, as a function of the degree exponent $\gamma$.
The original networks take $\kmin=2$ and $\kmax=100$ (the red dotted lines), $500$ (the black solid lines), and $2500$ (the blue dashed lines).
The lines in panel (a) (panel (b)) are drawn from Eq.~(\ref{eq:r0}) with Eqs.~(\ref{def:aBxy}) and~(\ref{def:aSx}) (with Eqs.~(\ref{def:bBxy}) and~(\ref{def:bSx})).
The black open circles represent the Monte Carlo results of $N=10^6$ and $\kmax=500$.
In the simulations, $r_0$ is averaged over $10^3$ network realizations.
}
\label{fig:assortativity}
\end{figure}

Figures~\ref{fig:assortativity}(a) and~\ref{fig:assortativity}(b) plot the assortativity coefficient $r_0$ as a function of the degree exponent $\gamma$ for scale-free networks with uniform reinforcement and selective reinforcement, respectively.
Here $\kmin=2$ and $\kmax$ takes $\kmax=100$, $500$ and $2500$.
The assortativity coefficient evaluated by Monte Carlo simulations with $\kmax=500$ is in good agreement with the theoretical line, which confirms the validity of our formulation.
Focusing on the region $\gamma>4$, we observe that the assortativity of the reinforced networks, either uniform or selective, is slightly positive.
Both the uniformly reinforced and selectively reinforced networks appear to have positive $r_0$ in the region where $\gamma$ is small; however, this is likely due to the finiteness of $\kmax$.
The figures show that $r$ approaches zero for $\gamma \le 4$ as $\kmax$ increases.




\begin{thebibliography}{40}
\expandafter\ifx\csname natexlab\endcsname\relax\def\natexlab#1{#1}\fi
\expandafter\ifx\csname bibnamefont\endcsname\relax
  \def\bibnamefont#1{#1}\fi
\expandafter\ifx\csname bibfnamefont\endcsname\relax
  \def\bibfnamefont#1{#1}\fi
\expandafter\ifx\csname citenamefont\endcsname\relax
  \def\citenamefont#1{#1}\fi
\expandafter\ifx\csname url\endcsname\relax
  \def\url#1{\texttt{#1}}\fi
\expandafter\ifx\csname urlprefix\endcsname\relax\def\urlprefix{URL }\fi
\providecommand{\bibinfo}[2]{#2}
\providecommand{\eprint}[2][]{\url{#2}}

\bibitem[{\citenamefont{Albert and Barab{\'a}si}(2002)}]{albert2002statistical}
\bibinfo{author}{\bibfnamefont{R.}~\bibnamefont{Albert}} \bibnamefont{and}
  \bibinfo{author}{\bibfnamefont{A.-L.} \bibnamefont{Barab{\'a}si}},
  \bibinfo{journal}{Reviews of Modern Physics} \textbf{\bibinfo{volume}{74}},
  \bibinfo{pages}{47} (\bibinfo{year}{2002}).

\bibitem[{\citenamefont{Newman}(2003{\natexlab{a}})}]{newman2003structure}
\bibinfo{author}{\bibfnamefont{M.~E.~J.} \bibnamefont{Newman}},
  \bibinfo{journal}{SIAM Review} \textbf{\bibinfo{volume}{45}},
  \bibinfo{pages}{167} (\bibinfo{year}{2003}{\natexlab{a}}).

\bibitem[{\citenamefont{Boccaletti et~al.}(2006)\citenamefont{Boccaletti,
  Latora, Moreno, Chavez, and Hwang}}]{boccaletti2006complex}
\bibinfo{author}{\bibfnamefont{S.}~\bibnamefont{Boccaletti}},
  \bibinfo{author}{\bibfnamefont{V.}~\bibnamefont{Latora}},
  \bibinfo{author}{\bibfnamefont{Y.}~\bibnamefont{Moreno}},
  \bibinfo{author}{\bibfnamefont{M.}~\bibnamefont{Chavez}}, \bibnamefont{and}
  \bibinfo{author}{\bibfnamefont{D.-U.} \bibnamefont{Hwang}},
  \bibinfo{journal}{Physics Reports} \textbf{\bibinfo{volume}{424}},
  \bibinfo{pages}{175} (\bibinfo{year}{2006}).

\bibitem[{\citenamefont{Albert et~al.}(2000)\citenamefont{Albert, Jeong, and
  Barab{\'a}si}}]{albert2000error}
\bibinfo{author}{\bibfnamefont{R.}~\bibnamefont{Albert}},
  \bibinfo{author}{\bibfnamefont{H.}~\bibnamefont{Jeong}}, \bibnamefont{and}
  \bibinfo{author}{\bibfnamefont{A.-L.} \bibnamefont{Barab{\'a}si}},
  \bibinfo{journal}{Nature} \textbf{\bibinfo{volume}{406}},
  \bibinfo{pages}{378} (\bibinfo{year}{2000}).

\bibitem[{\citenamefont{Callaway et~al.}(2000)\citenamefont{Callaway, Newman,
  Strogatz, and Watts}}]{callaway2000network}
\bibinfo{author}{\bibfnamefont{D.~S.} \bibnamefont{Callaway}},
  \bibinfo{author}{\bibfnamefont{M.~E.~J.} \bibnamefont{Newman}},
  \bibinfo{author}{\bibfnamefont{S.~H.} \bibnamefont{Strogatz}},
  \bibnamefont{and} \bibinfo{author}{\bibfnamefont{D.~J.} \bibnamefont{Watts}},
  \bibinfo{journal}{Physical Review Letters} \textbf{\bibinfo{volume}{85}},
  \bibinfo{pages}{5468} (\bibinfo{year}{2000}).

\bibitem[{\citenamefont{Cohen et~al.}(2000)\citenamefont{Cohen, Erez, ben
  Avraham, and Havlin}}]{cohen2000resilience}
\bibinfo{author}{\bibfnamefont{R.}~\bibnamefont{Cohen}},
  \bibinfo{author}{\bibfnamefont{K.}~\bibnamefont{Erez}},
  \bibinfo{author}{\bibfnamefont{D.}~\bibnamefont{ben Avraham}},
  \bibnamefont{and} \bibinfo{author}{\bibfnamefont{S.}~\bibnamefont{Havlin}},
  \bibinfo{journal}{Physical Review Letters} \textbf{\bibinfo{volume}{85}},
  \bibinfo{pages}{4626} (\bibinfo{year}{2000}).

\bibitem[{\citenamefont{Cohen et~al.}(2001)\citenamefont{Cohen, Erez, ben
  Avraham, and Havlin}}]{cohen2001breakdown}
\bibinfo{author}{\bibfnamefont{R.}~\bibnamefont{Cohen}},
  \bibinfo{author}{\bibfnamefont{K.}~\bibnamefont{Erez}},
  \bibinfo{author}{\bibfnamefont{D.}~\bibnamefont{ben Avraham}},
  \bibnamefont{and} \bibinfo{author}{\bibfnamefont{S.}~\bibnamefont{Havlin}},
  \bibinfo{journal}{Physical Review Letters} \textbf{\bibinfo{volume}{86}},
  \bibinfo{pages}{3682} (\bibinfo{year}{2001}).

\bibitem[{\citenamefont{Gross and Barth}(2022)}]{gross2022network}
\bibinfo{author}{\bibfnamefont{T.}~\bibnamefont{Gross}} \bibnamefont{and}
  \bibinfo{author}{\bibfnamefont{L.}~\bibnamefont{Barth}},
  \bibinfo{journal}{Frontiers in Physics} p. \bibinfo{pages}{561}
  (\bibinfo{year}{2022}).

\bibitem[{\citenamefont{Liu et~al.}(2005)\citenamefont{Liu, Wang, and
  Dang}}]{liu2005optimization}
\bibinfo{author}{\bibfnamefont{J.-G.} \bibnamefont{Liu}},
  \bibinfo{author}{\bibfnamefont{Z.-T.} \bibnamefont{Wang}}, \bibnamefont{and}
  \bibinfo{author}{\bibfnamefont{Y.-Z.} \bibnamefont{Dang}},
  \bibinfo{journal}{Modern Physics Letters B} \textbf{\bibinfo{volume}{19}},
  \bibinfo{pages}{785} (\bibinfo{year}{2005}).

\bibitem[{\citenamefont{Xiao et~al.}(2010)\citenamefont{Xiao, Xiao, Cheng, Ma,
  Fu, and Soh}}]{xiao2010robustness}
\bibinfo{author}{\bibfnamefont{S.}~\bibnamefont{Xiao}},
  \bibinfo{author}{\bibfnamefont{G.}~\bibnamefont{Xiao}},
  \bibinfo{author}{\bibfnamefont{T.}~\bibnamefont{Cheng}},
  \bibinfo{author}{\bibfnamefont{S.}~\bibnamefont{Ma}},
  \bibinfo{author}{\bibfnamefont{X.}~\bibnamefont{Fu}}, \bibnamefont{and}
  \bibinfo{author}{\bibfnamefont{H.}~\bibnamefont{Soh}}, \bibinfo{journal}{EPL
  (Europhysics Letters)} \textbf{\bibinfo{volume}{89}}, \bibinfo{pages}{38002}
  (\bibinfo{year}{2010}).

\bibitem[{\citenamefont{Schneider et~al.}(2011)\citenamefont{Schneider,
  Moreira, Andrade~Jr, Havlin, and Herrmann}}]{schneider2011mitigation}
\bibinfo{author}{\bibfnamefont{C.~M.} \bibnamefont{Schneider}},
  \bibinfo{author}{\bibfnamefont{A.~A.} \bibnamefont{Moreira}},
  \bibinfo{author}{\bibfnamefont{J.~S.} \bibnamefont{Andrade~Jr}},
  \bibinfo{author}{\bibfnamefont{S.}~\bibnamefont{Havlin}}, \bibnamefont{and}
  \bibinfo{author}{\bibfnamefont{H.~J.} \bibnamefont{Herrmann}},
  \bibinfo{journal}{Proceedings of the National Academy of Sciences}
  \textbf{\bibinfo{volume}{108}}, \bibinfo{pages}{3838} (\bibinfo{year}{2011}).

\bibitem[{\citenamefont{Wu and Holme}(2011)}]{wu2011onion}
\bibinfo{author}{\bibfnamefont{Z.-X.} \bibnamefont{Wu}} \bibnamefont{and}
  \bibinfo{author}{\bibfnamefont{P.}~\bibnamefont{Holme}},
  \bibinfo{journal}{Physical Review E} \textbf{\bibinfo{volume}{84}},
  \bibinfo{pages}{026106} (\bibinfo{year}{2011}).

\bibitem[{\citenamefont{Herrmann et~al.}(2011)\citenamefont{Herrmann,
  Schneider, Moreira, Andrade, and Havlin}}]{herrmann2011onion}
\bibinfo{author}{\bibfnamefont{H.~J.} \bibnamefont{Herrmann}},
  \bibinfo{author}{\bibfnamefont{C.~M.} \bibnamefont{Schneider}},
  \bibinfo{author}{\bibfnamefont{A.~A.} \bibnamefont{Moreira}},
  \bibinfo{author}{\bibfnamefont{J.~S.} \bibnamefont{Andrade}},
  \bibnamefont{and} \bibinfo{author}{\bibfnamefont{S.}~\bibnamefont{Havlin}},
  \bibinfo{journal}{Journal of Statistical Mechanics: Theory and Experiment}
  \textbf{\bibinfo{volume}{2011}}, \bibinfo{pages}{P01027}
  (\bibinfo{year}{2011}).

\bibitem[{\citenamefont{Zhou and Liu}(2014)}]{zhou2014memetic}
\bibinfo{author}{\bibfnamefont{M.}~\bibnamefont{Zhou}} \bibnamefont{and}
  \bibinfo{author}{\bibfnamefont{J.}~\bibnamefont{Liu}},
  \bibinfo{journal}{Physica A: Statistical Mechanics and its Applications}
  \textbf{\bibinfo{volume}{410}}, \bibinfo{pages}{131} (\bibinfo{year}{2014}).

\bibitem[{\citenamefont{Park and Hahn}(2016)}]{park2016bypass}
\bibinfo{author}{\bibfnamefont{J.}~\bibnamefont{Park}} \bibnamefont{and}
  \bibinfo{author}{\bibfnamefont{S.~G.} \bibnamefont{Hahn}},
  \bibinfo{journal}{Physical Review E} \textbf{\bibinfo{volume}{94}},
  \bibinfo{pages}{022310} (\bibinfo{year}{2016}).

\bibitem[{\citenamefont{Chujyo and Hayashi}(2021)}]{chujyo2021loop}
\bibinfo{author}{\bibfnamefont{M.}~\bibnamefont{Chujyo}} \bibnamefont{and}
  \bibinfo{author}{\bibfnamefont{Y.}~\bibnamefont{Hayashi}},
  \bibinfo{journal}{Applied Network Science} \textbf{\bibinfo{volume}{6}},
  \bibinfo{pages}{1} (\bibinfo{year}{2021}).

\bibitem[{\citenamefont{Lou et~al.}(2023)\citenamefont{Lou, Wang, and
  Chen}}]{lou2023structural}
\bibinfo{author}{\bibfnamefont{Y.}~\bibnamefont{Lou}},
  \bibinfo{author}{\bibfnamefont{L.}~\bibnamefont{Wang}}, \bibnamefont{and}
  \bibinfo{author}{\bibfnamefont{G.}~\bibnamefont{Chen}},
  \bibinfo{journal}{IEEE Circuits and Systems Magazine}
  \textbf{\bibinfo{volume}{23}}, \bibinfo{pages}{12} (\bibinfo{year}{2023}).

\bibitem[{\citenamefont{Beygelzimer et~al.}(2005)\citenamefont{Beygelzimer,
  Grinstein, Linsker, and Rish}}]{beygelzimer2005improving}
\bibinfo{author}{\bibfnamefont{A.}~\bibnamefont{Beygelzimer}},
  \bibinfo{author}{\bibfnamefont{G.}~\bibnamefont{Grinstein}},
  \bibinfo{author}{\bibfnamefont{R.}~\bibnamefont{Linsker}}, \bibnamefont{and}
  \bibinfo{author}{\bibfnamefont{I.}~\bibnamefont{Rish}},
  \bibinfo{journal}{Physica A: Statistical Mechanics and its Applications}
  \textbf{\bibinfo{volume}{357}}, \bibinfo{pages}{593} (\bibinfo{year}{2005}).

\bibitem[{\citenamefont{Zhao and Xu}(2009)}]{zhao2009enhancing}
\bibinfo{author}{\bibfnamefont{J.}~\bibnamefont{Zhao}} \bibnamefont{and}
  \bibinfo{author}{\bibfnamefont{K.}~\bibnamefont{Xu}},
  \bibinfo{journal}{Journal of Physics A: Mathematical and Theoretical}
  \textbf{\bibinfo{volume}{42}}, \bibinfo{pages}{195003}
  (\bibinfo{year}{2009}).

\bibitem[{\citenamefont{Li et~al.}(2012)\citenamefont{Li, Jia, Guan, and
  Wang}}]{li2012enhancing}
\bibinfo{author}{\bibfnamefont{L.}~\bibnamefont{Li}},
  \bibinfo{author}{\bibfnamefont{Q.-S.} \bibnamefont{Jia}},
  \bibinfo{author}{\bibfnamefont{X.}~\bibnamefont{Guan}}, \bibnamefont{and}
  \bibinfo{author}{\bibfnamefont{H.}~\bibnamefont{Wang}},
  \bibinfo{journal}{KSII Transactions on Internet and Information Systems
  (TIIS)} \textbf{\bibinfo{volume}{6}}, \bibinfo{pages}{1333}
  (\bibinfo{year}{2012}).

\bibitem[{\citenamefont{Ji et~al.}(2016)\citenamefont{Ji, Wang, Liu, Chen,
  Tang, Wei, and Tu}}]{ji2016improving}
\bibinfo{author}{\bibfnamefont{X.}~\bibnamefont{Ji}},
  \bibinfo{author}{\bibfnamefont{B.}~\bibnamefont{Wang}},
  \bibinfo{author}{\bibfnamefont{D.}~\bibnamefont{Liu}},
  \bibinfo{author}{\bibfnamefont{G.}~\bibnamefont{Chen}},
  \bibinfo{author}{\bibfnamefont{F.}~\bibnamefont{Tang}},
  \bibinfo{author}{\bibfnamefont{D.}~\bibnamefont{Wei}}, \bibnamefont{and}
  \bibinfo{author}{\bibfnamefont{L.}~\bibnamefont{Tu}},
  \bibinfo{journal}{Physica A: Statistical Mechanics and its Applications}
  \textbf{\bibinfo{volume}{444}}, \bibinfo{pages}{9} (\bibinfo{year}{2016}).

\bibitem[{\citenamefont{Chan and Akoglu}(2016)}]{chan2016optimizing}
\bibinfo{author}{\bibfnamefont{H.}~\bibnamefont{Chan}} \bibnamefont{and}
  \bibinfo{author}{\bibfnamefont{L.}~\bibnamefont{Akoglu}},
  \bibinfo{journal}{Data Mining and Knowledge Discovery}
  \textbf{\bibinfo{volume}{30}}, \bibinfo{pages}{1395} (\bibinfo{year}{2016}).

\bibitem[{\citenamefont{Cui et~al.}(2018)\citenamefont{Cui, Zhu, Wang, Xun, and
  Xia}}]{cui2018enhancing}
\bibinfo{author}{\bibfnamefont{P.}~\bibnamefont{Cui}},
  \bibinfo{author}{\bibfnamefont{P.}~\bibnamefont{Zhu}},
  \bibinfo{author}{\bibfnamefont{K.}~\bibnamefont{Wang}},
  \bibinfo{author}{\bibfnamefont{P.}~\bibnamefont{Xun}}, \bibnamefont{and}
  \bibinfo{author}{\bibfnamefont{Z.}~\bibnamefont{Xia}},
  \bibinfo{journal}{Physica A: Statistical Mechanics and its Applications}
  \textbf{\bibinfo{volume}{497}}, \bibinfo{pages}{185} (\bibinfo{year}{2018}).

\bibitem[{\citenamefont{Carchiolo et~al.}(2019)\citenamefont{Carchiolo,
  Grassia, Longheu, Malgeri, and Mangioni}}]{carchiolo2019network}
\bibinfo{author}{\bibfnamefont{V.}~\bibnamefont{Carchiolo}},
  \bibinfo{author}{\bibfnamefont{M.}~\bibnamefont{Grassia}},
  \bibinfo{author}{\bibfnamefont{A.}~\bibnamefont{Longheu}},
  \bibinfo{author}{\bibfnamefont{M.}~\bibnamefont{Malgeri}}, \bibnamefont{and}
  \bibinfo{author}{\bibfnamefont{G.}~\bibnamefont{Mangioni}},
  \bibinfo{journal}{Computational Social Networks}
  \textbf{\bibinfo{volume}{6}}, \bibinfo{pages}{1} (\bibinfo{year}{2019}).

\bibitem[{\citenamefont{Kazawa and Tsugawa}(2020)}]{kazawa2020effectiveness}
\bibinfo{author}{\bibfnamefont{Y.}~\bibnamefont{Kazawa}} \bibnamefont{and}
  \bibinfo{author}{\bibfnamefont{S.}~\bibnamefont{Tsugawa}},
  \bibinfo{journal}{Physica A: Statistical Mechanics and its Applications}
  \textbf{\bibinfo{volume}{545}}, \bibinfo{pages}{123586}
  (\bibinfo{year}{2020}).

\bibitem[{\citenamefont{Dong et~al.}(2020)\citenamefont{Dong, Tian, Tang, Li,
  and Lai}}]{dong2020improving}
\bibinfo{author}{\bibfnamefont{Z.}~\bibnamefont{Dong}},
  \bibinfo{author}{\bibfnamefont{M.}~\bibnamefont{Tian}},
  \bibinfo{author}{\bibfnamefont{R.}~\bibnamefont{Tang}},
  \bibinfo{author}{\bibfnamefont{X.}~\bibnamefont{Li}}, \bibnamefont{and}
  \bibinfo{author}{\bibfnamefont{J.}~\bibnamefont{Lai}},
  \bibinfo{journal}{Nonlinear Dynamics} \textbf{\bibinfo{volume}{100}},
  \bibinfo{pages}{2287} (\bibinfo{year}{2020}).

\bibitem[{\citenamefont{Chen et~al.}(2022)\citenamefont{Chen, Zhao, Qin, Meng,
  and Gao}}]{chen2022robustness}
\bibinfo{author}{\bibfnamefont{C.-Y.} \bibnamefont{Chen}},
  \bibinfo{author}{\bibfnamefont{Y.}~\bibnamefont{Zhao}},
  \bibinfo{author}{\bibfnamefont{H.}~\bibnamefont{Qin}},
  \bibinfo{author}{\bibfnamefont{X.}~\bibnamefont{Meng}}, \bibnamefont{and}
  \bibinfo{author}{\bibfnamefont{J.}~\bibnamefont{Gao}},
  \bibinfo{journal}{Physica A: Statistical Mechanics and its Applications}
  \textbf{\bibinfo{volume}{604}}, \bibinfo{pages}{127851}
  (\bibinfo{year}{2022}).

\bibitem[{\citenamefont{Chujyo and Hayashi}(2022)}]{chujyo2022adding}
\bibinfo{author}{\bibfnamefont{M.}~\bibnamefont{Chujyo}} \bibnamefont{and}
  \bibinfo{author}{\bibfnamefont{Y.}~\bibnamefont{Hayashi}},
  \bibinfo{journal}{Plos One} \textbf{\bibinfo{volume}{17}},
  \bibinfo{pages}{e0276733} (\bibinfo{year}{2022}).

\bibitem[{\citenamefont{Zhuo et~al.}(2011)\citenamefont{Zhuo, Peng, Liu, Liu,
  and Long}}]{zhuo2011improving}
\bibinfo{author}{\bibfnamefont{Y.}~\bibnamefont{Zhuo}},
  \bibinfo{author}{\bibfnamefont{Y.}~\bibnamefont{Peng}},
  \bibinfo{author}{\bibfnamefont{C.}~\bibnamefont{Liu}},
  \bibinfo{author}{\bibfnamefont{Y.}~\bibnamefont{Liu}}, \bibnamefont{and}
  \bibinfo{author}{\bibfnamefont{K.}~\bibnamefont{Long}},
  \bibinfo{journal}{Physica Scripta} \textbf{\bibinfo{volume}{83}},
  \bibinfo{pages}{025801} (\bibinfo{year}{2011}).

\bibitem[{\citenamefont{Wang et~al.}(2021)\citenamefont{Wang, Zeng, Liu, and
  Chen}}]{wang2021adversarial}
\bibinfo{author}{\bibfnamefont{C.}~\bibnamefont{Wang}},
  \bibinfo{author}{\bibfnamefont{C.}~\bibnamefont{Zeng}},
  \bibinfo{author}{\bibfnamefont{H.}~\bibnamefont{Liu}}, \bibnamefont{and}
  \bibinfo{author}{\bibfnamefont{J.}~\bibnamefont{Chen}},
  \bibinfo{journal}{Electronics} \textbf{\bibinfo{volume}{10}},
  \bibinfo{pages}{2614} (\bibinfo{year}{2021}).

\bibitem[{\citenamefont{Newman and Ziff}(2000)}]{newman2000efficient}
\bibinfo{author}{\bibfnamefont{M.~E.~J.} \bibnamefont{Newman}}
  \bibnamefont{and} \bibinfo{author}{\bibfnamefont{R.~M.} \bibnamefont{Ziff}},
  \bibinfo{journal}{Physical Review Letters} \textbf{\bibinfo{volume}{85}},
  \bibinfo{pages}{4104} (\bibinfo{year}{2000}).

\bibitem[{\citenamefont{Newman and Ziff}(2001)}]{newman2001fast}
\bibinfo{author}{\bibfnamefont{M.~E.~J.} \bibnamefont{Newman}}
  \bibnamefont{and} \bibinfo{author}{\bibfnamefont{R.~M.} \bibnamefont{Ziff}},
  \bibinfo{journal}{Physical Review E} \textbf{\bibinfo{volume}{64}},
  \bibinfo{pages}{016706} (\bibinfo{year}{2001}).

\bibitem[{\citenamefont{Albert et~al.}(1999)\citenamefont{Albert, Jeong, and
  Barab{\'a}si}}]{albert1999diameter}
\bibinfo{author}{\bibfnamefont{R.}~\bibnamefont{Albert}},
  \bibinfo{author}{\bibfnamefont{H.}~\bibnamefont{Jeong}}, \bibnamefont{and}
  \bibinfo{author}{\bibfnamefont{A.-L.} \bibnamefont{Barab{\'a}si}},
  \bibinfo{journal}{Nature} \textbf{\bibinfo{volume}{401}},
  \bibinfo{pages}{130} (\bibinfo{year}{1999}).

\bibitem[{\citenamefont{Leskovec and Krevl}(2014)}]{snapnets}
\bibinfo{author}{\bibfnamefont{J.}~\bibnamefont{Leskovec}} \bibnamefont{and}
  \bibinfo{author}{\bibfnamefont{A.}~\bibnamefont{Krevl}},
  \emph{\bibinfo{title}{{SNAP Datasets}: {Stanford} large network dataset
  collection}}, \bibinfo{howpublished}{\url{http://snap.stanford.edu/data}}
  (\bibinfo{year}{2014}).

\bibitem[{\citenamefont{Leskovec et~al.}(2005)\citenamefont{Leskovec,
  Kleinberg, and Faloutsos}}]{leskovec2005graphs}
\bibinfo{author}{\bibfnamefont{J.}~\bibnamefont{Leskovec}},
  \bibinfo{author}{\bibfnamefont{J.}~\bibnamefont{Kleinberg}},
  \bibnamefont{and}
  \bibinfo{author}{\bibfnamefont{C.}~\bibnamefont{Faloutsos}}, in
  \emph{\bibinfo{booktitle}{Proceedings of the eleventh ACM SIGKDD
  international conference on Knowledge discovery in data mining}}
  (\bibinfo{year}{2005}), pp. \bibinfo{pages}{177--187}.

\bibitem[{\citenamefont{Newman}(2002)}]{newman2002assortative}
\bibinfo{author}{\bibfnamefont{M.~E.~J.} \bibnamefont{Newman}},
  \bibinfo{journal}{Physical Review Letters} \textbf{\bibinfo{volume}{89}},
  \bibinfo{pages}{208701} (\bibinfo{year}{2002}).

\bibitem[{\citenamefont{Newman}(2003{\natexlab{b}})}]{newman2003mixing}
\bibinfo{author}{\bibfnamefont{M.~E.~J.} \bibnamefont{Newman}},
  \bibinfo{journal}{Physical Review E} \textbf{\bibinfo{volume}{67}},
  \bibinfo{pages}{026126} (\bibinfo{year}{2003}{\natexlab{b}}).

\bibitem[{\citenamefont{Shiraki and Kabashima}(2010)}]{shiraki2010cavity}
\bibinfo{author}{\bibfnamefont{Y.}~\bibnamefont{Shiraki}} \bibnamefont{and}
  \bibinfo{author}{\bibfnamefont{Y.}~\bibnamefont{Kabashima}},
  \bibinfo{journal}{Physical Review E} \textbf{\bibinfo{volume}{82}},
  \bibinfo{pages}{036101} (\bibinfo{year}{2010}).

\bibitem[{\citenamefont{Tanizawa et~al.}(2012)\citenamefont{Tanizawa, Havlin,
  and Stanley}}]{tanizawa2012robustness}
\bibinfo{author}{\bibfnamefont{T.}~\bibnamefont{Tanizawa}},
  \bibinfo{author}{\bibfnamefont{S.}~\bibnamefont{Havlin}}, \bibnamefont{and}
  \bibinfo{author}{\bibfnamefont{H.~E.} \bibnamefont{Stanley}},
  \bibinfo{journal}{Physical Review E} \textbf{\bibinfo{volume}{85}},
  \bibinfo{pages}{046109} (\bibinfo{year}{2012}).

\bibitem[{\citenamefont{Mizutaka and Tanizawa}(2016)}]{mizutaka2016robustness}
\bibinfo{author}{\bibfnamefont{S.}~\bibnamefont{Mizutaka}} \bibnamefont{and}
  \bibinfo{author}{\bibfnamefont{T.}~\bibnamefont{Tanizawa}},
  \bibinfo{journal}{Physical Review E} \textbf{\bibinfo{volume}{94}},
  \bibinfo{pages}{022308} (\bibinfo{year}{2016}).

\end{thebibliography}
\end{document}